\documentclass[a4paper,twocolumn,aps,10pt,longbibliography]{revtex4-1}
\usepackage{amsmath}
\usepackage{amssymb}
\usepackage{graphicx}
\usepackage{amsthm}

\let\<\langle
\let\>\rangle
\let\l\left
\let\r\right
\def\({\left(}
\def\){\right)}
\def\lll{\left\{}
\def\rrr{\right\}}
\def\llll{\left[}
\def\rrrr{\right]}
\def\tr{\mathrm{Tr}}

\newtheorem*{assumption}{Assumption}
\newtheorem*{theorem}{Theorem}
\newtheorem{lemma}{Lemma}

\def\ASSUMPTIONb{\begin{assumption}}
\def\ASSUMPTIONe{\end{assumption}}
\def\THEOREMb{\begin{theorem}}
\def\THEOREMe{\end{theorem}}
\def\LEMMAb{\begin{lemma}}
\def\LEMMAe{\end{lemma}}
\def\PROOFb{\begin{proof}}
\def\PROOFe{\end{proof}}

\begin{document}

\title{Realistic Area-Law Bound on Entanglement from Exponentially Decaying Correlations}
\author{Jaeyoon Cho}
\affiliation{Asia Pacific Center for Theoretical Physics, Pohang 37673, Korea}
\affiliation{Department of Physics, POSTECH, Pohang 37673, Korea}
\date{\today}
\begin{abstract}
A remarkable feature of typical ground states of strongly-correlated many-body systems is that the entanglement entropy is not an extensive quantity. In one dimension, there exists a proof that a finite correlation length sets a constant upper-bound on the entanglement entropy, called the area law. However, the known bound exists only in a hypothetical limit, rendering its physical relevance highly questionable. In this paper, we give a simple proof of the area law for entanglement entropy in one dimension under the condition of exponentially decaying correlations. Our proof dramatically reduces the previously known bound on the entanglement entropy, bringing it, for the first time, into a realistic regime. The proof is composed of several simple and straightforward steps based on elementary quantum information tools. We discuss the underlying physical picture, based on a renormalization-like construction underpinning the proof, which transforms the entanglement entropy of a continuous region into a sum of mutual informations in different length scales and the entanglement entropy at the boundary.
\end{abstract}
\maketitle

\section{Introduction}

Understanding the universal nature of strongly-correlated many-body systems is one of the central topics in theoretical physics. Even though strongly-correlated systems are generally intractable, it is possible to unfold the universal relationship between their characteristic attributes, providing a guiding principle for studying specific model Hamiltonians. For example, the existence or absence of a spectral gap, a finite or diverging correlation length, and the behavior of entanglement entropies are commonly studied attributes, which find intriguing mutual connections~\cite{eis10, has06, nac06, gos16, has07, aha11, ara12, ara13, deb10, cho14, has07b, bra13, bra14}.

One of the prominent open problems in this context is whether the ground states of gapped Hamiltonians always obey the area law for entanglement entropy in any dimension, i.e., whether the entanglement between a subregion and its complement scales as the boundary size of the chosen region or can grow faster, e.g., as the volume of the region~\cite{eis10}. The underlying idea is that the existence of a gap significantly restricts the correlation that the ground state can accommodate. There is a well-established theorem, namely, the exponential-clustering theorem, which states that the existence of a spectral gap implies a finite correlation length in the ground state~\cite{has06, nac06, gos16}. Indeed, a seminal work by Hastings~\cite{has07} and several ensuing works~\cite{aha11, ara12, ara13} have given proofs of the area law in one-dimensional gapped systems, wherein the area law means a constant bound on the entanglement entropy. In higher dimensional cases, however, only partial results are present~\cite{deb10, cho14}. Originally spawned by the Bekenstein-Hawking entropy~\cite{bek73, haw74}, the area law has arrested a huge interest over the last decade, thanks to its widespread relevance, e.g., to frameworks based on tensor network states~\cite{sch11, oru14}, topological entanglement entropies~\cite{kit06, lev06}, the holographic formula based on the AdS/CFT correspondence~\cite{ryu06}, the Hamiltonian complexity theory~\cite{gha14}, and so on.

Since the proof for one-dimensional gapped systems, a naturally ensuing question was whether a finite correlation length alone can imply the area law. Albeit likely at first glance, serious doubt was cast upon its possibility due to unfavorable examples such as quantum data-hiding states and quantum expander states, for which a small correlation and a large entanglement can coexist~\cite{hay04, has07b, bra13, bra14}. Amid such uncertainty, the recent proof that a finite correlation length indeed implies the area law in one dimension was a remarkable achievement~\cite{bra13, bra14}. However, the physical relevance of that proof is highly questionable because the obtained upper-bound of the entanglement entropy is ridiculously huge to such an extent that it is never reachable in any physically sensible situation (having a constant of $\sim 10^8$ in the exponent, the bound easily surpasses the estimated number of atoms in the whole universe!)~\cite{bra14}. Consequently, we are still facing a quite unsatisfactory situation: under the condition of a finite correlation length alone, does the upper-bound of entanglement entropies exist only in such a hypothetical limit? Answering this question is important in truly confirming our picture on one-dimensional systems: in one dimension, a finite gap implies exponential decay of correlations, which in turn implies the area law. Here, the aforementioned unfavorable examples again seem to suggest that this picture might be misleading in reality.

In this paper, we give a proof of the one-dimensional entanglement area law from exponentially decaying correlations, which dramatically reduces the previously obtained bound and, for the first time, brings the bound into a realistic regime. As well as the involved constants, our bound also improves the asymptotic dependence on the correlation length. With $\xi$ being the correlation length, we obtain the bound of $\sim ( \log\xi ) 2^{ \text{(const.)} \xi } $, while the previous proof gives $\sim \xi^{ \text{(const.)} \xi } $~\cite{bra13, bra14}. 
In fact, there are a number of strong cases for the exponential dependence on the correlation length being unavoidable (see Ref.~\cite{has16} for the current state of understanding), which suggests that our bound leaves only little room for improvement.
Interestingly, the dependence on $\xi$ is even more favorable than that of Hastings' original proof for gapped systems, which reads $\sim \xi ( \log\xi ) 2^{ \text{(const.)} \xi } $~\cite{has07}, although this bound was significantly improved by a recent work (for gapped systems)~\cite{ara13}. 

\begin{figure}
\includegraphics[width=0.98\columnwidth]{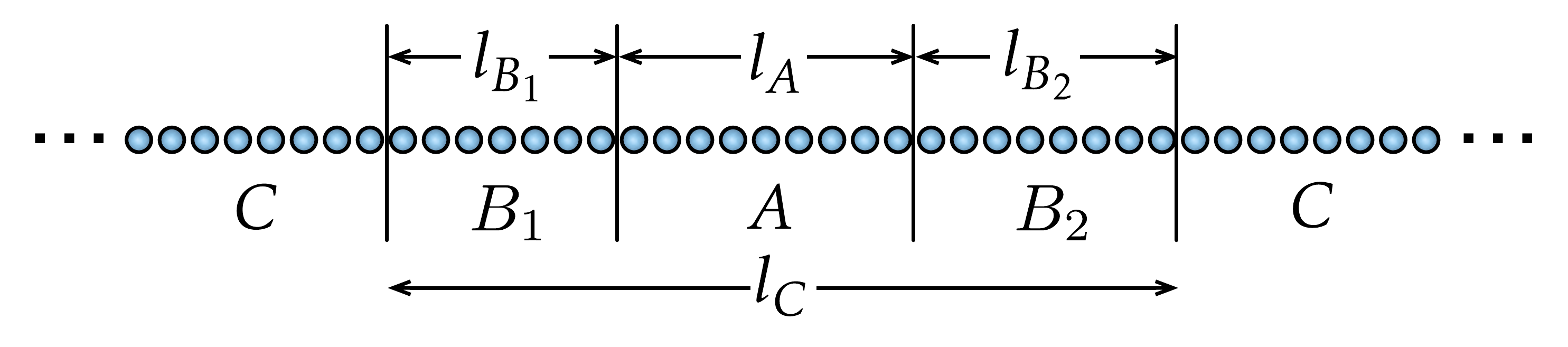}
\caption{The basic partitioning of the system for the proof. Here, $l_{B_1}\le l_{B_2}$ as a convention. We let $l_{b} = l_{B_1}$ and $l_B = l_{B_1} + l_{B_2}$. $l_C$ is defined to be $l_A+l_B$ because $S(C)=S(AB)$.}
\label{fig:1}
\end{figure}

Moreover, compared to the previous one, our proof is remarkably simpler and more straightforward. The proof directly addresses the internal structure of the states with exponentially decaying correlations using elementary quantum information tools. Such a direct nature allows us to envisages a clear and intuitive picture on the encountered situation. The central part of the proof is to show that when the length scale is increased as $\ell_0 \rightarrow x\ell_0 \rightarrow x^2 \ell_0 \rightarrow \cdots \rightarrow x^n \ell_0$ with $x>0$, the upper-bound of the mutual information $I(A:C)$ in Fig.~\ref{fig:1} initially increases indefinitely, but saturates at some point, and then decreases exponentially in $n$. Combined with a simple renormalization-like construction, this behavior of the mutual information accounts not only for the entanglement area law, but also for why the area-law bound is exponentially large in the correlation length and how the common intuition---with a finite correlation length, the entanglement of a large region is determined by the correlations around the boundary---indeed makes sense.

Thus, the present work makes our view on one-dimensional systems quite solid and consistent. We hope that our proof offers a more direct and detailed insight into the situation and becomes an important step towards the understanding of the area law in higher dimensions. 

\section{Main theorem}

We consider a one-dimensional chain of qubits, i.e., $s=1/2$ spins, in a pure state $\rho=|\Psi\>\<\Psi|$. This setting covers the cases of higher dimensional spins because one can then decompose each spin into a set of qubits and rescale the length. We make the following assumption:
\ASSUMPTIONb
For arbitrary operators $X$ and $Y$ supported, respectively, on regions $R_{X}$ and $R_{Y}$ separated by a graph distance $l$, the following inequality holds:
\begin{equation}
	\l| \<X \otimes Y\> - \<X\>\<Y\> \r| \le \|X\| \, \|Y\| 2^{-l/\xi},
	\label{eq:assum}
\end{equation}
where $\xi$ is the correlation length of the system, $\|\cdot\|$ denotes the operator norm, and $\<\cdot\> = \tr(\cdot\,\rho)$. Without loss of generality, we assume $\xi \ge 1$.
\ASSUMPTIONe

Under the above assumption, we prove the following theorem.
\THEOREMb
For any real parameter $\alpha_0 \in [2/3,1)$, the entanglement entropy $S$ of an arbitrary continuous region is bounded by
\begin{equation}
	S < \frac{ \alpha_0 }{ 1 - \alpha_0 } \( \log \frac{ \xi }{ 1 - \alpha_0 } + 3 \) 4^{ n_0 } + 12,
	\label{eq:bound}
\end{equation}
where
\begin{equation}
	n_0 \le \l\lceil \frac{ 10\xi }{ \alpha_0 } + \frac{ 1 - \alpha_0 }{ \alpha_0 } 
	     	 \frac{ 3 }{ \log \xi - \log ( 1 - \alpha_0 ) + 3  } \r\rceil + 2
\end{equation}
with $\lceil x \rceil$ denoting the smallest integer larger than or equal to $x$.
\THEOREMe
For example, if $\alpha_0 = 10 / 11$ is taken, we obtain
\begin{equation}
	S < 160 \( \log \xi + 6.5 \) 4^{ \lceil 11\xi + 0.05 \rceil } + 12.
\end{equation}
We note that the theorem is obtained by a particular choice of the parameters used in the proof. Thus, the constants in the theorem do not represent the optimal ones. 
Having said that, it is nevertheless unlikely that the constants could be significantly reduced, unless the proof is enhanced by a completely new idea.

We also note that the main assumption~\eqref{eq:assum} can be generalized to some extent.
See Sec.~\ref{sec:discussion} for the details.

\section{Overall Picture}
\label{sec:iii}

Let us partition the chain into regions $A$, $B=B_1+B_2$, and $C$, as shown in Fig.~\ref{fig:1}. This partitioning is our basic setting throughout the proof. Our plan is to inspect the entropic relations between the subregions while changing their sizes. The size of each region is denoted by $l$ with the corresponding subscript as in the caption of the figure. We will denote the reduced density matrix by $\rho$ with the corresponding superscript, e.g., $\rho^C = \tr_{AB}\rho$, and use the similar convention for other density matrices. The von Neumann entropy of a local region is denoted, e.g., by $S(C)=S(\rho^C)$, etc.

To begin with, recall the definition of the mutual information:
\begin{equation}
	\begin{split}
		I(A:C) & = S(A) + S(C) - S(AC) \ge 0,\\
		I(B_1:B_2) & = S(B_1) + S(B_2) - S(B) \ge 0.
	\end{split}
\end{equation}
One can rearrange the terms using $S(C) = S(B_1 A B_2)$ and $S(AC) = S(B)$, ending up with a nice structure:
\begin{equation}
	S(B_1 A B_2) = S(B_1) - S(A) + S(B_2) + f(B_1: A: B_2), 
	\label{eq:11212222}
\end{equation}
where
\begin{equation}
	f(B_1:A:B_2) \equiv I(A:C) - I(B_1:B_2)
\end{equation}
is defined for later convenience.

\begin{figure*}
    \includegraphics[width=0.98\textwidth]{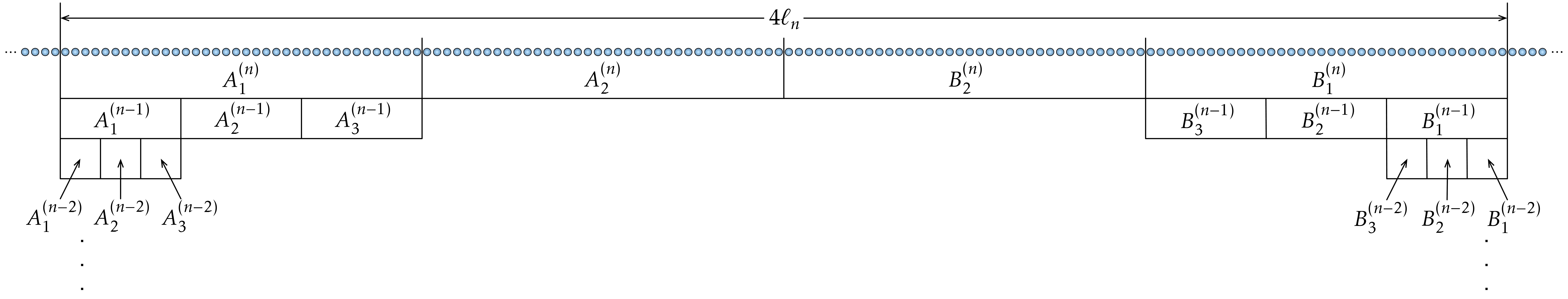}
    \caption{Partitioning of the system for the proof of Lemma~\ref{lem:7}. The region of interest, which is of size $4\ell_n$, is composed of four regions of size $\ell_n$. We recursively decompose the two end regions $A_1^{(i)}$ and $B_1^{(i)}$, respectively, into three regions of size $\ell_{i-1}$ following the rule: $A_1^{(i)} = A_1^{(i-1)} + A_2^{(i-1)} + A_3^{(i-1)}$ and $B_1^{(i)} = B_3^{(i-1)} + B_2^{(i-1)} + B_1^{(i-1)}$. Each region denoted by superscript $(i)$ is of size $\ell_i = 3^i \ell_0$.}
    \label{fig:2}
\end{figure*}

The expression~\eqref{eq:11212222} can be utilized as follows. Consider the partitioning in Fig.~\ref{fig:2}. Let $\ell_n = 3^n \ell_0$ with some small unit length $\ell_0$ and integer $n$. We are interested in the entanglement entropy of large region $A_1^{(n)}A_2^{(n)}B_2^{(n)}B_1^{(n)}$ of length $4\ell_n$. Eq.~\eqref{eq:11212222} can be applied to this partitioning in two different ways: 
\begin{widetext}
\begin{equation}
\begin{split}
	S ( A_1^{(n)} A_2^{(n)} B_2^{(n)} B_1^{(n)} ) 
	& = S ( A_1^{(n)} A_2^{(n)} ) - S ( B_2^{(n)} ) + S ( B_1^{(n)} ) 
		+ f ( A_1^{(n)} A_2^{(n)} : B_2^{(n)} : B_1^{(n)} ), \\
	S ( A_1^{(n)} A_2^{(n)} B_2^{(n)} B_1^{(n)} ) 
	& = S ( A_1^{(n)} ) - S( A_2^{(n)} ) + S ( B_2^{(n)} B_1^{(n)} ) 
		+ f ( A_1^{(n)} : A_2^{(n)} : B_2^{(n)} B_1^{(n)} ).
\label{eq:71}
\end{split}
\end{equation}
Performing a similar task to the end regions,
\begin{equation}
\begin{split}
	S ( A_1^{(i)} A_2^{(i)} ) 
	& = S ( A_1^{(i-1)} A_2^{(i-1)} ) - S ( A_3^{(i-1)} ) + S ( A_2^{(i)} ) 
		+ f ( A_1^{(i-1)} A_2^{(i-1)}: A_3^{(i-1)}: A_2^{(i)} ), \\
	S ( A_1^{(i)} ) 
	& = S ( A_1^{(i-1)} ) - S ( A_2^{(i-1)} ) + S ( A_3^{(i-1)} ) 
		+ f ( A_1^{(i-1)}: A_2^{(i-1)}: A_3^{(i-1)} ), \\
   	S ( B_1^{(i)} ) 
	& = S ( B_3^{(i-1)} ) - S ( B_2^{(i-1)} ) + S ( B_1^{(i-1)} ) 
		+ f ( B_3^{(i-1)}: B_2^{(i-1)}: B_1^{(i-1)} ), \\
	S ( B_2^{(i)} B_1^{(i)} ) 
	& = S ( B_2^{(i)} ) - S ( B_3^{(i-1)} ) + S ( B_2^{(i-1)} B_1^{(i-1)} ) 
		+ f ( B_2^{(i)}: B_3^{(i-1)}: B_2^{(i-1)} B_1^{(i-1)} ).
\label{eq:72}
\end{split}
\end{equation}
Summing Eqs.~\eqref{eq:71} and \eqref{eq:72} over $1 \le i \le n$, most of the $S(\cdot)$ terms are cancelled out. The result is
\begin{equation}
\begin{split}
	S ( A_1^{(n)} A_2^{(n)} B_2^{(n)} B_1^{(n)} ) 
	& = \frac12 \l\{ S ( A_1^{(0)} A_2^{(0)} ) + S ( A_1^{(0)} ) 
	      - S ( A_2^{(0)} ) 
		  - S ( B_2^{(0)} ) + S ( B_1^{(0)} ) + S ( B_2^{(0)} B_1^{(0)} ) 
	  + f(\cdots) \r\} \\
    & = S ( A_1^{(0)} ) + S ( B_1^{(0)} ) 
		  - \frac12 \l\{ I(A_1^{(0)}:A_2^{(0)}) + I(B_2^{(0)}:B_1^{(0)})\r\}+\frac12 f(\cdots),
\end{split}
\label{eq:11221211}
\end{equation}
\end{widetext}
where $f(\cdots)$ is the abbreviation of the sum of all the mutual information terms 
in Eqs.~\eqref{eq:71} and \eqref{eq:72}
with different length scales. Note here that all the mutual information terms corresponding to $I(A:C)$ in Fig.~\ref{fig:1} are added, whereas those corresponding to $I(B_1:B_2)$ are always subtracted.

The expression~\eqref{eq:11221211} reduces the problem into figuring out how the mutual information scales with varying the length scale.
We prove in the next section that if $\{l_{B_1},l_A,l_{B_2}\}$ in Fig.~\ref{fig:1} are all comparable to $x^n \ell_0$ with $x>0$, the behavior of the upper-bound of $I(A:C)$ with respect to increasing $n$ is such that (i) it initially increases indefinitely for $n \lesssim n_0$ with $n_0$ being linearly large in the correlation length $\xi$ (transient behavior), but (ii) it saturates around $n \simeq n_0$ (saturation), and then (iii) it decreases exponentially in $n$ for $n \gtrsim n_0$ (asymptotic behavior). A similar behavior can be proven for $I(B_1:B_2)$ by a slight modification of the proof, albeit not to be shown explicitly. Together with the expression~\eqref{eq:11221211}, such behavior of the mutual information produces three straightforward implications. 

First, the entanglement entropy of an arbitrarily large region of length $4\ell_n$ is upper-bounded by a constant
as $f(\cdots)$ converges due to the asymptotic behavior of the mutual information.
This leads to the entanglement area law as follows. For an arbitrary continuous region of length $l$, one can choose the partitioning in Fig.~\ref{fig:1} such that $l_A=l$ and $l_{B_1} = l_{B_2} = 4\ell_n - l$ with sufficiently large $n$. Then, from the strong subadditivity $S(A) \le S(B_1 A)+S(AB_2) - S(C)$, we find the entanglement entropy of an arbitrary region is upper-bounded by twice the maximum entanglement entropy of a region of length $4\ell_n$. This bound is the inequality~\eqref{eq:bound} in the main theorem.

Second, the upper-bound of the entanglement entropy is mostly determined by the transient and saturation behavior. As $n_0$ is linear in $\xi$, the area-law bound is exponentially large in the correlation length. 

Third, Eq.~\eqref{eq:11221211} is valid for any choice of $\ell_0$. 
Our plan is to take the size of $A_1^{(0)}$ and $B_1^{(0)}$ to be the saturation length scale, which will be denoted by $l_0$, instead of a small unit length $\ell_0$.
Then, the entanglement entropy is bounded by the entanglement entropies of the two boundary regions $A_1^{(0)}$ and $B_1^{(0)}$, where the remaining mutual information terms in $f(\cdots)$ are treated as a finite correction to the bound. 
The point to be addressed here is that if we change $l_0$ to $3^m l_0$ with positive integer $m$, then while the size of the boundary region is increased, the sum of the remaining mutual information terms decays exponentially in $m$ due to the aforementioned asymptotic behavior. 
The physical meaning of this is that if the correlation length is finite, the entanglement entropy of a large region is determined by the correlations around the boundary, which is consistent with our common intuition. 
A caveat is that as mentioned above, the boundary region responsible for the entanglement entropy is exponentially large in the correlation length in general.

\section{Existence of an Area-Law Bound}

The state of the system can be written in a Schmidt-decomposed form as
\begin{equation}
	|\Psi\> = \sum_{i=1}^M \sqrt{p_i} |i\>^A \otimes |\phi_i\>^{BC}, \quad 
	p_1 \ge p_2 \ge \cdots \ge p_M,
	\label{eq:state}
\end{equation}
where the Schmidt coefficients $p_i$'s are sorted in descending order and $M \le 2^{l_A}$ is the Schmidt number.
If a local projection 
\begin{equation}
	P_{mn} \equiv \sum_{i=m}^n |i\>^A \<i|
\end{equation}
on region $A$ is applied, the resulting normalized state of region $C$ is given by 
\begin{equation}
	\rho_{mn}^C \equiv \sum_{i=m}^n \frac{p_i}{Q_{mn}} \phi_{ii}^C,
\end{equation}
where we define 
\begin{equation}
	\begin{split}
		Q_{mn} & \equiv \< P_{mn} \> = \sum_{i=m}^n p_i,\\
		\phi_{ij}^C & \equiv \tr_B |\phi_i\>^{BC}\<\phi_j|.
	\end{split}
\end{equation}
The exponential decay of correlations implies that the larger $Q_{mn}$ and $l_b$ are, the closer $\rho_{mn}^C$ is to the original state $\rho^C$ of region $C$:

\LEMMAb
	\begin{equation}
		\mathcal{D} (\rho^C, \rho_{mn}^C) \le \frac{2^{-l_b/\xi}}{Q_{mn}},
	\end{equation}
	where
	\begin{equation}
		\mathcal{D} (\rho, \sigma) \equiv \frac{1}{2} \| \rho - \sigma \|_1 
		= \max_{0 \le \Lambda \le I} \tr \{ \Lambda (\rho - \sigma) \}
	\end{equation}
	denotes the trace distance between two states $\rho$ and $\sigma$.
	\label{lem:1}
\LEMMAe

\PROOFb
	For any operator $\Lambda^C$ supported on region $C$,
	\begin{equation}
		\begin{split}
		\l| \tr \{ \Lambda^C (\rho^C - \rho_{mn}^C) \} \r| 
		& = \l| \l< \(I - \frac{P_{mn}}{Q_{mn}}\) \otimes \Lambda^C \r> \r|\\
		& \le \l\| I - \frac{P_{mn}}{Q_{mn}} \r\| \| \Lambda^C \| \, 2^{-l_b/\xi},
		\end{split}
	\end{equation}
	where we use the assumption~\eqref{eq:assum} and $\< I - P_{mn} / Q_{mn} \> = 0$.
	By choosing $0 \le \Lambda^C \le I$ maximizing the trace, we obtain the lemma.
\PROOFe

An important step is to define $q(\alpha) \in \{ 0, 1, \cdots, M \}$ with a control parameter $\alpha \in (0, 1)$ to be chosen later in such a way that the number of $p_i$'s with $p_i \ge 2^{-\alpha l_b / \xi}$ is $q(\alpha)$.
As $p_i$'s are sorted in descending order, $q(\alpha)$ plays the role of a {\em cut-off} index in that $p_{q(\alpha)} \ge 2^{-\alpha l_b / \xi}$ and $p_{q(\alpha)+1} < 2^{-\alpha l_b / \xi}$ unless $q(\alpha)=0$ or $M$.
For $i \le q(\alpha)$, each $\phi_{ii}^C$ alone is a good approximation to $\rho^C$ as $\mathcal{D} (\rho^C, \rho_{ii}^C) \le 2^{-(1-\alpha)l_b / \xi}$ from Lemma~\ref{lem:1}.
For $i > q(\alpha)$, on the other hand, individual $\phi_{ii}^C$'s are basically {\em uncertain} and should be considered arbitrary apart from the constraint in Lemma~\ref{lem:1}.
This uncertain portion amounts to
\begin{equation}
	Q(\alpha) \equiv \sum_{p_i < 2^{-\alpha l_b / \xi}} p_i.
	\label{eq:Q}
\end{equation}
For $q(\alpha) < M$, $Q(\alpha) = Q_{q(\alpha)+1, M}$ and if $q(\alpha) = M$, $Q(\alpha) = 0$.
Note that $Q(\alpha)$ is a function of region $A$ and $l_b$ in Fig.~\ref{fig:1}.
For brevity, we simply denote $Q(\alpha)$, the meaning of which will be clear from the context.
We will later see that when $\{l_{B_1}, l_A, l_{B_2}\}$ are all comparable, $Q(\alpha)$ asymptotically vanishes for a large length scale, which is an important part of the proof.

The Fannes' inequality states that for states $\rho$ and $\sigma$ acting on a Hilbert space of dimension $d$, if $\mathcal{D}(\rho, \sigma) = \epsilon \in [0,1]$, $| S(\rho) - S(\sigma) | \le \epsilon \log d - \epsilon \log \epsilon - (1-\epsilon) \log (1-\epsilon)$~\cite{nie00}.
If $\epsilon \in [0, 1/2]$, we can use a modified form
\begin{equation}
	|S(\rho) - S(\sigma)| \le \epsilon (\log d - 2 \log \epsilon)
	\label{eq:fineq}
\end{equation}
as $\epsilon \log \epsilon - (1-\epsilon) \log (1-\epsilon) \le -2\epsilon \log \epsilon$.
From this modified Fannes' inequality, Lemma~\ref{lem:1} directly leads to the following lemma.

\LEMMAb
	If $Q_{mn} \ge 2^{-\alpha l_b /\xi}$ and $(1-\alpha) l_b \ge \xi$, then
	\begin{equation}
		| S(C) - S(\rho_{mn}^C) | \le \epsilon( l_B + \log M, l_b, \alpha ),
	\end{equation}
	where
	\begin{equation}
		\epsilon( L, l_b, \alpha ) \equiv \{ L + 2(1 - \alpha)l_b \} 2^{-(1-\alpha)l_b/\xi}.
		\label{eq:epsilon}
	\end{equation}
	\label{lem:2}
\LEMMAe

\PROOFb
	For $Q_{mn} \ge 2^{-\alpha l_b / \xi}$, $\mathcal{D} (\rho^C, \rho_{mn}^C) \le 2^{-(1-\alpha)l_b/\xi}$ from Lemma~\ref{lem:1}.
	As the Hilbert-space dimension for region $C$ is upper-bounded by $M2^{l_B}$ from the expression~\eqref{eq:state}, the Fannes' inequality~\eqref{eq:fineq} leads to the lemma.
\PROOFe

Note that $\log M \le l_A$.
If $\{l_{B_1}, l_A, l_{B_2}\}$ are all comparable to each other, the bound in Lemma~\ref{lem:2} decreases exponentially in $l_b$. 

Let us define another important quantity, namely, $S(l) \in [0,l]$, which denotes the maximum entropy of a continuous region of length $l$.
Furthermore, let us define the maximum entropy per site with a slight modification:
\begin{equation}
	\bar s(l) \equiv \frac{ S(l) + \epsilon_h }{ l },
	\label{eq:262157}
\end{equation}
where 
\begin{equation}
	\epsilon_h \equiv \max_{ p \in [0,1] } (-p \log p) < 0.531.
\end{equation}
The additional constant is added merely for simplicity of the ensuing formulae.
Note that from the subadditivity of entropy,
\begin{equation}
	\bar s(nl) \le \bar s(l) \quad \text{for any positive integer $n$.}
	\label{eq:sub}
\end{equation}
It turns out that $Q(\alpha)$ is related to $\bar s(l)$ from the concavity of the entropy for region $A$.

\LEMMAb
	\begin{equation}
		Q(\alpha) \le \frac{\xi}{\alpha l_b}\{S(A) + \epsilon_h\}
		\le \frac{\xi l_A}{\alpha l_b} \bar s(l_A).
	\end{equation}
	\label{lem:3}
\LEMMAe

\PROOFb
	If $q(\alpha) = M$, the lemma is satisfied as $Q(\alpha) = 0$.
	Suppose $q(\alpha) < M$.
	Given various possible cases of $\{ p_i \}$,
	$S( \rho_{mM}^A ) = - \sum_{i=m}^M (p_i / Q_{mM}) \log (p_i / Q_{mM})$ is minimal when $p_i = p_m$ for $m \le i < M$ and $p_M \le p_m$.
	Thus,
	\begin{equation*}
		\begin{split}
			& S(\rho_{mM}^A) \\
			& \ge \frac{Q_{mM} - p_M}{p_m} \( -\frac{p_m}{Q_{mM}} \log \frac{p_m}{Q_{mM}} \)
			  - \frac{p_M}{Q_{mM}} \log \frac{p_M}{Q_{mM}} \\
			& = -\frac{Q_{mM} - p_M}{Q_{mM}} \log p_m - \frac{p_M}{Q_{mM}} \log p_M + \log Q_{mM} \\
			& \ge -\log p_m + \log Q_{mM}.
		\end{split}
	\end{equation*}
	From this and the concavity of entropy,
	\begin{equation*}
		\begin{split}
			S(A) 
			& \ge Q(\alpha) S(\rho_{q(\alpha)+1,M}^A) \\
			& \ge Q(\alpha) \( \frac{\alpha l_b}{\xi} + \log Q(\alpha) \)
			  \ge Q(\alpha) \frac{\alpha l_b}{\xi} - \epsilon_h,
		\end{split}
	\end{equation*}
	which implies the lemma.	
\PROOFe

In addition, the concavity of the entropy for region $B$ results in the following lemma.

\LEMMAb
	If $(1-\alpha)l_b \ge \xi$, then
	\begin{equation}
		S(C) \le \frac{1}{1-Q(\alpha)} S(B) + \epsilon(l_C, l_b, \alpha).
	\end{equation}
	\label{lem:4}
\LEMMAe

\PROOFb
	If $q(\alpha)=0$, the lemma is satisfied as $Q(\alpha) = 1$.
	Suppose $q(\alpha) > 0$.
	From the concavity of entropy,
	\begin{gather}
		S(B) \ge \{ 1 - Q(\alpha) \} S(\rho_{1q(\alpha)}^B), \\
		S(\rho_{1q(\alpha)}^B)
		\ge \sum_{i=1}^{q(\alpha)} \frac{p_i}{1-Q(\alpha)} S(\phi_{ii}^B).
	\end{gather}
	The second inequality implies that there exists a certain $i_0 \in \{1, 2, \cdots, q(\alpha)\}$ such that $S(\phi_{i_0 i_0}^B) \le S(\rho_{1q(\alpha)}^B)$.
	The lemma then follows because $S(\phi_{i_0 i_0}^B) = S(\phi_{i_0 i_0}^C)$ and $S(C) \le S(\phi_{i_0 i_0}^C) + \epsilon(l_C, l_b, \alpha)$ from Lemma~\ref{lem:2}.	
\PROOFe

Suppose $l_{B_1}:l_A:l_{B_2}=1:(x-2):1$ for some integer $x \ge 3$ and $l_b = l_0$,
hence $l_C = xl_0$. Note $S(B) \le 2S(l_0)$ from the subadditivity of entropy.
If $l_0$ is sufficiently large so that $\epsilon(l_C, l_b, \alpha)  \le \epsilon_h$,
Lemma~\ref{lem:4} implies
\begin{equation}
	\bar s(xl_0) 
	\le \frac{1}{1 - Q(\alpha)} \frac{2}{x} \bar s(l_0).
	\label{eq:20}
\end{equation}
On the other hand, as $\bar s(l_A) = \bar s((x-2)l_0) \le \bar s(l_0)$ from the subadditivity~\eqref{eq:sub}, $Q(\alpha)$ is also bounded by $\bar s(l_0)$ from Lemma~\ref{lem:3}.
The inequality~\eqref{eq:20} thus turns into
\begin{equation}
	\bar s(xl_0) \le \frac{1}{1 - \{(x-2)\xi/\alpha\} \bar s(l_0)} \frac{2}{x} \bar s(l_0)
	\equiv \gamma(x,l_0) \bar s(l_0),
	\label{eq:21}
\end{equation}
where we define a function $\gamma(x,l)$ accordingly.
Consequently, if there exists $l_0$ with sufficiently small $\bar s(l_0)$ so that $\gamma(x,l_0) < 1$, then
\begin{equation}
	\bar s(x^n l_0) \le \lll \prod_{i=0}^{n-1} \gamma(x, x^i l_0) \rrr \bar s(l_0)
	< \gamma_0^n \bar s(l_0),
	\label{eq:22}
\end{equation}
where $\gamma_0 \equiv \gamma(x, l_0)$. Note that $\lim_{n\rightarrow\infty} \gamma(x, x^n l_0) = 2/x$ if $\gamma_0 < 1$.

The inequality~\eqref{eq:22} and Lemma~\ref{lem:3} indicate that both $\bar s(x^n l_0)$ and $Q(\alpha)$ asymptotically decay exponentially in $n$. It turns out later that this is related to the asymptotic behavior of the mutual information mentioned in the previous section. 
Before proceeding, however, we need to make sure that there indeed exists such $l_0$ that makes $\bar s(l_0)$ sufficiently small.
Only then, the above argument is valid.
The inequality~\eqref{eq:20} is insufficient here as $Q(\alpha)$ may be arbitrarily close to one in the first place.

In order to obtain a complementary inequality, we revisit the subadditivity of entropy, which can be derived from the expression~\eqref{eq:state}:
\begin{equation}
	\begin{split}
		S(AB) 
		& = S(C) = S \( \sum_{i=1}^M p_i \phi_{ii}^C \) \\
		& \le H(\{p_i\}) + \sum_{i=1}^M p_i S(\phi_{ii}^C) \\
		& = H(\{p_i\}) + \sum_{i=1}^M p_i S(\phi_{ii}^B) \\
		& \le S(A) + S(B),
	\end{split}
	\label{eq:24}
\end{equation}
where
\begin{equation}
 	H(\{p_i\}) \equiv -\sum_i p_i \log p_i
\end{equation}
denotes the Shannon entropy.
The second line is a general inequality~\cite{nie00}, the third line uses $S(\phi_{ii}^C) = S(\phi_{ii}^B)$, and the last line uses the concavity of entropy.
The following lemma is obtained by incorporating Lemma~\ref{lem:2} into the inequality~\eqref{eq:24}.

\LEMMAb
	If $Q(\alpha) \ge 2^{-\alpha l_b/\xi}$ and $(1-\alpha)l_b \ge \xi$, then
	\begin{equation}
		S(C) \le S(A) + S(B) - Q(\alpha)\frac{\alpha l_b}{\xi} + 2\epsilon_h + \epsilon(l_C, l_b, \alpha).
	\end{equation}
	\label{lem:5}
\LEMMAe

\PROOFb
Let us slightly modify the inequality~\eqref{eq:24}.
That is, let us group M $p_i$'s into $K \le M$ sets as $\{R_1, R_2, \cdots, R_K\}$ with $R_m = Q_{i_m,i_{m+1}-1}$, where $i_1=1 < i_2 < \cdots < i_{K+1} = M+1$.
We then have
\begin{equation}
	\begin{split}
		H(\{p_i\}) 
		& = \sum_{m=1}^K R_m \sum_{i=i_m}^{i_{m+1}-1} \frac{p_i}{R_m} 
		    \( - \log \frac{p_i}{R_m} - \log R_m\) \\
		& = -\sum_{m=1}^K R_m \sum_{i=i_m}^{i_{m+1}-1} \frac{p_i}{R_m} \log \frac{p_i}{R_m}
		    + H(\{ R_m \})
	\end{split}
	\label{eq:25}
\end{equation}
and
\begin{equation}
	\sum_{i=1}^M p_i S(\phi_{ii}^C) = \sum_{m=1}^K R_m \sum_{i=i_m}^{i_{m+1}-1} \frac{p_i}{R_m} S(\phi_{ii}^C).
	\label{eq:26}
\end{equation}
But, again from the inequality
\begin{equation}
	\begin{split}
		& S(\rho_{i_m,i_{m+1}-1}^C) =  S \( \sum_{i=i_m}^{i_{m+1}-1} \frac{p_i}{R_m} \phi_{ii}^C \) \\
		& \le - \sum_{i=i_m}^{i_{m+1}-1} \frac{p_i}{R_m} \log \frac{p_i}{R_m}
		+ \sum_{i=i_m}^{i_{m+1}-1} \frac{p_i}{R_m} S( \phi_{ii}^C ),
	\end{split}
\end{equation}
the inequality~\eqref{eq:24}, along with Eqs.~\eqref{eq:25} and \eqref{eq:26}, becomes
\begin{equation}
	\sum_{m=1}^K R_m S(\rho_{i_m,i_{m+1}-1}^C) \le S(A) + S(B) - H(\{R_m\}).
	\label{eq:29}
\end{equation}
Let us now actually group the indices so that
\begin{equation*}
    \begin{cases}
        i_m = i 
        & \text{for } 1 \le m \le q(\alpha), \\
        2^{-\alpha l_b / \xi} \le R_m \le 2 \cdot 2^{-\alpha l_b / \xi}
        & \text{for } m > q(\alpha),
    \end{cases}
\end{equation*}
which is always possible as $p_i < 2^{-\alpha l_b / \xi}$ for $i > q(\alpha)$ and $Q(\alpha) \ge 2^{-\alpha l_b / \xi}$. 
Note $R_m \ge 2^{-\alpha l_b / \xi}$ for all $m$.
Then, from Lemma~\ref{lem:2}, $S(C) \le S(\rho_{i_m,i_{m+1}-1}^C) + \epsilon(l_C, l_b, \alpha)$ for all $m$ and hence the inequality~\eqref{eq:29} becomes
\begin{equation}
	S(C) \le S(A) + S(B) - H(\{R_m\}) + \epsilon(l_C, l_b, \alpha).
\end{equation}
To obtain the lower bound of $H(\{R_m\})$, we follow the same logic as in the proof of Lemma~\ref{lem:3} and use the concavity of entropy to find out
\begin{equation}
	\begin{split}
    H(\{R_m\}) 
    & \ge Q(\alpha) \lll - \log \( 2 \cdot 2^{-\alpha l_b / \xi} \) 
                                    + \log Q(\alpha) \rrr \\
    & \ge Q(\alpha) \frac{\alpha l_b}{\xi} 
                     + \min_{x\in[0,1]} \( - x + x \log x \).
	\end{split}
\end{equation}
As $\min_x (-x + x \log x) = \min_x \{-2x + 2x \log (2x)\} = -2 \epsilon_h$,
\begin{equation}
    H(\{R_m\}) \ge Q(\alpha) \frac{\alpha l_b}{\xi} - 2 \epsilon_h,
\end{equation}
which implies the lemma.
\PROOFe

Suppose $l_{B_1}:l_A:l_{B_2}=1:(x-2):1$ and $l_b = \ell_0$.
If $x\ge4$ and $\ell_0$ is sufficiently large so that $\epsilon(l_C, l_b, \alpha) \le \epsilon_h$, Lemma~\ref{lem:5} implies
\begin{equation}
	\bar s(x\ell_0) \le \bar s(\ell_0) - Q(\alpha) \frac{\alpha}{x\xi},
	\label{eq:34}
\end{equation}
where we have used the subadditivity~\eqref{eq:sub}.
It can be seen that the inequalities~\eqref{eq:20} and \eqref{eq:34} can play complementary roles.
Note that for any $Q_c\in (0,1)$, if $Q(\alpha) \le Q_c$, $\bar s(xl) \le \frac{2\bar s(l)}{x(1-Q_c)}$ from the inequality~\eqref{eq:20}, and if $Q(\alpha) \ge Q_c$, $\bar s(xl) \le \bar s(l)-Q_c \frac{\alpha}{x\xi}$ from the inequality~\eqref{eq:34}. We thus find
\begin{equation}
	\bar s(xl) \le \max \lll \frac{2\bar s(l)}{x(1-Q_c)},\bar s(l)-Q_c\frac{\alpha}{x\xi}\rrr.
	\label{eq:240204}
\end{equation}
For example, take $x=4$ and $Q_c=2/5$. 
Then, $\bar s(xl) \le \bar s(l)-\alpha/10\xi$ if $\bar s(l) \ge 3\alpha / 5\xi$ because in this case the second term on the right-hand side is larger than or equal to the first term.
Consequently, if we assume $\bar s(x^r \ell_0) \ge 3\alpha/5\xi$ for all positive integer $r$, 
$0 \le \bar s(x^r \ell_0) \le \bar s(\ell_0) - r\alpha/10\xi$ leads to a contradiction, which means there exists $r \lesssim 10\xi/\alpha$ such that $\bar s(x^r \ell_0) < 3\alpha/5\xi$. 
Once we reach this point, the inequality~\eqref{eq:240204} implies $\bar s(x^{r+1}\ell_0) < (5/6)\bar s(x^r \ell_0)$ because now the first term on the right-hand side is larger than the second term.
In the same manner, for positive integer $m$, $\bar s(x^{r+m}\ell_0) < (5/6)^m (3\alpha/5\xi)$, which decreases exponentially in $m$ (in fact, the decrement is much faster because $Q_c$ can be adjusted optimally at each step).
We thus come to a conclusion that for arbitrarily small $\epsilon>0$, there exists $l_0=x^{n_0}\ell_0$ such that $\bar s(l_0) < \epsilon$ with $n_0$ being linearly large in $\xi$ and $\log(1/\epsilon)$, hence the inequality~\eqref{eq:22} is indeed valid.
The detailed calculation is presented in the next section.

The next step is to see how mutual information $I(A:C)$ scales with increasing the length scale as $l_0 \rightarrow xl_0 \rightarrow x^2l_0 \rightarrow \cdots$.
As the mutual information quantifies the amount of information shared between two regions, the exponentially decaying correlation is expected to imply a decay of $I(A:C)$ with increasing the length scale.
The explicit bound of $I(A:C)$ can be obtained by separating the state $\rho^{AC} = \sum_{i,j=1}^M \sqrt{p_i p_j} |i\>^A\<j|\otimes \phi_{ij}^C$ into three parts as
\begin{equation*}
	\rho^{AC} = \left[ \sum_{i,j=1}^{q(\alpha)} 
				+ \sum_{i,j=q(\alpha)+1}^M
				+ \sum_\text{the rest}\right] \sqrt{p_i p_j} |i\>^A\<j|\otimes \phi_{ij}^C
\end{equation*}
and inspecting how each part contributes to $S(AC)$.
From a rather straightforward analysis, we end up with the following lemma.

\LEMMAb
	For $\alpha\in(0,1/2)$, if $(1-2\alpha)l_b \ge \xi$, then
	\begin{equation}
		\begin{split}
			I(A:C) 
			& \le 2Q(\alpha) S(C) + 2 H(\{Q(\alpha), 1-Q(\alpha)\}) \\
			& \quad + 2\epsilon(l_C, l_b, \alpha) + \epsilon(l_B+l_b, l_b, 2\alpha).
		\end{split}
		\label{eq:36}
	\end{equation}
	If $Q(\alpha) < 1/2$,
	\begin{equation}
		\begin{split}
			I(A:C) 
			& \le 2Q(\alpha) S(C) - 4Q(\alpha) \log Q(\alpha) \\
			& \quad + 2\epsilon(l_C, l_b, \alpha) + \epsilon(l_B+l_b, l_b, 2\alpha).
		\end{split}
		\label{eq:92258}
	\end{equation}
	\label{lem:6}
\LEMMAe
The detailed proof is given in the Appendix.

Suppose $\{l_{B_1}, l_A, l_{B_2}\}$ are small-integer multiples of $x^n l_0$.
Then, $Q(\alpha)S(C) \le \mathcal{O}[\bar{s}(l_A)\bar{s}(l_C)l_C] \le \mathcal{O}[\bar{s}(x^n l_0)^2 l_C] \le \mathcal{O}[(\gamma_0^2 x)^n]$ from Lemma~\ref{lem:3}, the subadditivity~\eqref{eq:sub}, and the inequality~\eqref{eq:22}, where $\mathcal{O}[\cdot]$ omits the constant prefactor.
Similarly, $-Q(\alpha)\log Q(\alpha) \le \mathcal{O}[n\gamma_0^n]+\mathcal{O}[\gamma_0^n]$ from Lemma~\ref{lem:3}.
Consequently, if $\gamma_0<1/\sqrt{x}$, the right-hand side of the inequality~\eqref{eq:36} is exponentially small in $n$.
This is possible with $x>4$ and sufficiently small $\bar{s}(l_0)$ as can be seen from the inequality~\eqref{eq:21}.

We are now fully ready for the final stage of the proof. Let us define
\begin{equation}
	\eta(l_1,l_2,l_3) \equiv \max_{l_{B_1}=l_1,l_A=l_2,l_{B_2}=l_3} I(A:C).
\end{equation}
Then, from the expression~\eqref{eq:11221211}, the following lemma follows.

\LEMMAb
	For $l_n=3^n l_0$,
	\begin{equation}
			S(4l_n) \le 2 S(l_0) + \eta(l_n,l_n,2l_n) + \sum_{0\le i<n} \lambda_i,
			\label{eq:92354}
	\end{equation}
	where $\lambda_i \equiv \eta(2l_i, l_i, 3l_i) + \eta(l_i,l_i,l_i)$.
	\label{lem:7}
\LEMMAe

This lemma states that $S(4l_n)$ is upper-bounded by a constant $2S(l_0)$ independent of $n$, up to the correction terms.
The remaining task is to ensure that the correction terms are indeed upper-bounded by a constant.
This is easily done in view of our analysis above.
We have shown above that 
\begin{equation}
	\begin{split}
	I(A:C) & \le \mathcal{O}[(\gamma_0^2 x)^n] + \mathcal{O}[n\gamma_0^n] + \mathcal{O}[\gamma_0^n] \\
	&\quad + \mathcal{O}[l_b 2^{-(1-\alpha)l_b/\xi}] + \mathcal{O}[l_b 2^{-(1-2\alpha)l_b/\xi}]
	\end{split}
	\label{eq:292359}
\end{equation}
when $\{l_{B_1}, l_{A}, l_{B_2}\}$ are small-integer multiples of $x^n l_0$ with $x>4$, where it is understood that $\bar{s}(l_0)$ can be made arbitrarily small so that $\gamma_0^2 x<1$.
This directly implies that $\lambda_{2m}$ decays exponentially in $m$ as $l_{2m}=9^m l_0$ (i.e., $x=9$), hence $\sum_{m=0}^\infty \lambda_{2m}$ is a finite constant.
Likewise, $\lambda_{2m+1}$ also decays exponentially in $m$, hence $\sum_{m=0}^\infty \lambda_{2m+1}$ is a finite constant.
As a result, $\sum_{i=0}^\infty \lambda_i$ is upper-bounded by a constant, as we anticipated.

The analytical bound can be obtained by summing the right-hand side of the inequality~\eqref{eq:292359} over $n$ and restoring the omitted constant prefactors according to Lemma~\ref{lem:7}.
For example, from the inequality~\eqref{eq:92258} and Lemma~\ref{lem:3},
\begin{equation}
\begin{split}
	\eta(2l_{2i}, l_{2i}, 3l_{2i}) 
	& < 2\frac{\xi}{2\alpha}\bar s(l_{2i}) \bar s(6l_{2i}) 6l_{2i} \\
	& \quad -4\frac{\xi}{2\alpha}\bar s(l_{2i}) \log \lll \frac{\xi}{2\alpha}\bar s(l_{2i})\rrr,
\end{split}
\end{equation}
where we use the fact that $\epsilon(\cdots)$ terms are much smaller than $2\frac{\xi}{2\alpha}\bar s(l_{2i})6l_{2i}\epsilon_h$ that appears out of the definition \eqref{eq:262157}.
Now suppose we get $\bar s(l_0) < \beta/\xi$ from the inequality~\eqref{eq:240204}.
As explained, one can always find such $l_0$ for arbitrarily small $\beta$.
We then find, using the subadditivity $\bar s(6l_{2i}) < \bar s(l_{2i})$ and the inequality~\eqref{eq:22},
\begin{equation}
	\eta(2l_{2i}, l_{2i}, 3l_{2i}) 
	< 6\frac{\beta}{\alpha}(9\gamma_0^2)^i l_0
	-2\frac{\beta}{\alpha}\gamma_0^i \log \lll \frac{\beta}{2\alpha}\gamma_0^i\rrr.
	\label{eq:151125}
\end{equation}
Apparently, $\sum_{i=0}^\infty \eta(2l_{2i}, l_{2i}, 3l_{2i})$ converges if $9\gamma_0^2 < 1$, which is again the case for sufficiently small $\bar s(l_0)$.
In the same manner, one can bound $\eta(2l_{2i+1},l_{2i+1},3l_{2i+1})$, $\eta(l_{2i},l_{2i},l_{2i})$, and $\eta(l_{2i+1},l_{2i+1},l_{2i+1})$, and their summations over $i$, which completes the proof.

Note that in obtaining the inequality~\eqref{eq:151125}, we applied the inequality~\eqref{eq:22} with $\bar s(x^n l_0) < \gamma_0^n \bar s(l_0)$ instead of $\bar s(x^n l_0) \le \prod_{i=0}^{n-1} \gamma(9,9^i l_0) \bar s(l_0)$ that is tighter.
This was simply for brevity of the presentation.
By replacing $\gamma_0^i$ in the inequality~\eqref{eq:151125} with $\prod_{j=0}^{i-1} \gamma(9,9^j l_0)$, one obtains a much tighter bound.
As the inequality~\eqref{eq:21} is nonlinear, this tighter bound should be obtained with the aid of numerical summation.
In the next section, we present the details of the calculation.

\section{Calculation of the area-law bound}

As explained in the previous section, the proof is composed of two main steps. 
In the first step, we use Lemmas~\ref{lem:4} and \ref{lem:5} to prove that there exists a certain $l_0$ such that $\bar s(l_0)$ is below a small threshold.
In the second step, we use Lemmas~\ref{lem:3}, \ref{lem:4}, \ref{lem:6}, and \ref{lem:7} to upper-bound $S(4 \cdot 3^n l_0)$.
The area-law bound is then $S(l) \le 2S(4 \cdot 3^n l_0)$ for arbitrary $l$ from the strong subadditivity of entropy.

\subsection{First step}

The following lemma incorporates Lemmas~\ref{lem:4} and \ref{lem:5}.
\LEMMAb
Fix $\alpha_0 \in [2/3, 1)$ and let $\ell_n = 4^n \ell_0$ and $\bar s_n = \bar s( \ell_n )$ with
\begin{equation}
	\ell_0 = \frac{2\xi}{1 - \alpha_0} \( \log \frac{\xi}{1 - \alpha_0} + 3 \).
	\label{eq:42}
\end{equation}
Then, for any $Q_c \in [0,1]$,
\begin{equation}
	\begin{cases}
	\displaystyle\bar s_{n+1} < \bar s_n - Q_c \frac{\alpha_0}{4\xi} 
	& \displaystyle\text{if }\bar s_n \ge \frac{Q_c ( 1 - Q_c ) \alpha_0}{2 ( 1 - 2 Q_c ) \xi} > 0,\\
	\displaystyle\bar s_{n+1} < \frac{\bar s_n}{2 ( 1 - Q_c )}
	& \displaystyle\text{if }\bar s_n \le \frac{Q_c ( 1 - Q_c ) \alpha_0}{2 ( 1 - 2 Q_c ) \xi}.\\
	\end{cases}
	\label{eq:43}
\end{equation}
\label{lem:8}
\LEMMAe

\begin{proof}
Suppose $l_{B_1} : l_A : l_{B_2} = 1:2:1$ and $l_b = \ell_n$, hence $l_C = \ell_{n+1}$.
Denote $Q( \alpha_0 )$ in this setting by $Q_n$. 
Let $g = 8\xi / (1-\alpha_0)$. 
Then, $\ell_0 = (g/4)\log g$.
For this choice of $\ell_0$, 
$\epsilon(l_C, l_b, \alpha_0) \le \epsilon(4\ell_0,\ell_0,\alpha_0) \le (7g/6)(\log g)2^{-2\log g} = (7/6)(\log g)/g < \epsilon_h$ as $g \ge 24$.
Lemmas~\ref{lem:4} and \ref{lem:5} then imply
\begin{equation}
\begin{split}
	\bar s_{n+1} & < \frac{\bar s_n}{2 ( 1 - Q_n )}, \\
	\bar s_{n+1} & < \bar s_n - Q_n \frac{\alpha_0}{4\xi}.	
	\label{eq:81}
\end{split}
\end{equation}
Note that for arbitrary $Q_c$, if $Q_n \le Q_c$, $\bar s_{n+1} < \frac{\bar s_n}{2 ( 1 - Q_c )}$, and if $Q_{n} \ge Q_{c}$, $\bar s_{n+1} < \bar s_{n} - Q_{c} \frac{\alpha_{0}}{4\xi}$. 
Thus,
\begin{equation}
	\bar s_{n+1} < \max \lll \frac{\bar s_{n}}{2 ( 1 - Q_{c} )} ,
							   \bar s_{n} - Q_{c} \frac{\alpha_{0}}{4\xi} \rrr.
\end{equation}
If $\bar s_{n} \ge \frac{Q (1-Q) \alpha_{0}}{2 (1-2Q) \xi} > 0$, $\frac{\bar s_{n}}{2 (1-Q)} \le \bar s_{n} - Q \frac{\alpha_{0}}{4\xi}$ and if $\bar s_{n} \le \frac{Q (1-Q) \alpha_{0}}{2 (1-2Q) \xi}$, $\frac{\bar s_{n}}{2 (1-Q)} \ge \bar s_{n} - Q \frac{\alpha_{0}}{4\xi}$. This observation leads to the lemma.
\end{proof}

We are now in a position to find $l_0$ with a sufficiently small $\bar s(l_0)$. 
Our target is $\bar s ( l_0 ) < \alpha_0 / 27\xi$. 

\LEMMAb
	Inherit the definitions in Lemma~\ref{lem:8}. For any $\alpha_0 \in [2/3, 1)$, there exists $l_0$ and $n_0$ such that
	\begin{equation}
		\begin{split}
			l_0 & \le 4^{ n_0 } \ell_0 \\
			n_0 & \le \l\lceil \frac{10\xi}{\alpha_{0}} \( 1 + \frac{\epsilon_{h}}{\ell_{0}}\) \r\rceil + 2, \\
			\bar s ( l_0 ) & < \frac{ \alpha_0 }{ 27\xi },
		\end{split}
	\end{equation}
	which implies
	\begin{equation}
		S ( l_0 ) < \frac{ 2 \alpha_0 }{ 27 ( 1 - \alpha_0 ) }
	    	        \l( \log \frac { \xi }{ 1 - \alpha_0 } + 3 \r) 4^{ n_0 }.
	\end{equation}
	\label{lem:9}
\LEMMAe

\begin{proof}
Let us use Lemma~\ref{lem:8} with $Q_c = 2/5$, for which
\begin{equation*}
	\frac{Q_c(1-Q_c)\alpha_0}{2(1-2Q_c)\xi} = \frac{3\alpha_0}{5\xi}.
\end{equation*}
Let $r=\lceil (10\xi/\alpha)(1+\epsilon_h/\ell_0) \rceil$.
If we assume $\bar s_n \ge 3 \alpha_{0} /5 \xi$ for all integer $n \le r - 6$, then from the first inequality in Lemma~\ref{lem:8},
\begin{equation*}
	\frac{3\alpha_{0}}{5\xi} \le \bar s_{r - 6} < \bar s_0 - (r - 6) \frac{\alpha_0}{10\xi} = \bar s_0 - \frac{\alpha_0}{10\xi} r + \frac{3\alpha_0}{5\xi}
\end{equation*}
leads to a contradiction because this means
\begin{equation*}
	\frac{\alpha_0}{10\xi} \l\lceil \frac{10\xi}{\alpha_0}\( 1 + \frac{\epsilon_h}{\ell_0} \) \r\rceil
	= \frac{\alpha_0}{10\xi} r
	< \bar s_0 \le 1 + \frac{\epsilon_h}{\ell_0}.
\end{equation*}
We thus come to a conclusion that there exists an integer $n_0' \le r - 6$ such that $\bar s_{n_0'} < 3 \alpha_{0} / 5 \xi$.

Once we have arrived at this, we use the second inequality in Lemma~\ref{lem:8} recursively with a proper choice of $Q_{c}$. 
That is, as $\bar s_{n_0'} < {3\alpha_0}/{5\xi}$, we find $Q_c$ that satisfies $\frac{Q_c ( 1 - Q_c ) \alpha_0}{2 ( 1 - 2 Q_c ) \xi} = \frac{3\alpha_0}{5\xi}$, which is $Q_c=2/5$, so that the second inequality in Lemma~\ref{lem:8} is applicable.
Then, the inequality indicates $\bar s_{n_0' + 1} < {\alpha_0}/{2\xi}$.
Now we find $Q_c$ that satisfies $\frac{Q_c ( 1 - Q_c ) \alpha_0}{2 ( 1 - 2 Q_c ) \xi} = \frac{\alpha_0}{2\xi}$, which is $Q_c=(3-\sqrt{5})/2\simeq 0.38$, from which $\bar s_{n_0'+2} < \alpha_0 /2(\sqrt{5}-1)\xi \simeq 0.40\alpha_0/\xi$.
We repeat this procedure as follows.
For brevity, we present only numerical values up to the second significant digit: 
$\bar s_{n_0' + 2} < 0.41 {\alpha_0}/{\xi}$
$\rightarrow$
$Q_{c} = 0.36$
$\rightarrow$
$\bar s_{n_0' + 3} < 0.32 {\alpha_0}/{\xi}$
$\rightarrow$
$Q_{c} = 0.33$
$\rightarrow$
$\bar s_{n_0' + 4} < 0.24 {\alpha_0}/{\xi}$
$\rightarrow$
$Q_{c} = 0.28$
$\rightarrow$
$\bar s_{n_0' + 5} < 0.17 {\alpha_{0}}/{\xi}$
$\rightarrow$
$Q_{c} = 0.23$
$\rightarrow$
$\bar s_{n_0' + 6} < 0.11 {\alpha_{0}}/{\xi}$
$\rightarrow$
$Q_{c} = 0.17$
$\rightarrow$
$\bar s_{n_0' + 7} < 0.064 {\alpha_{0}}/{\xi}$
$\rightarrow$
$Q_{c} = 0.11$
$\rightarrow$
$\bar s_{n_0' + 8} < 0.036 {\alpha_{0}}/{\xi}$
$< {\alpha_{0}}/{27\xi}$.
This implies the lemma.
\end{proof}

\subsection{Second step}

The remaining task is to obtain the upper-bound of the summation in Lemma~\ref{lem:7}.
Let $l_n \equiv 3^n l_0$.
As explained in the previous section, it is convenient to decompose the summation as $\sum_i \lambda_i = \sum_m \lambda_{2m} + \sum_m \lambda_{2m+1}$.

First, we need to obtain the upper-bounds of $\{\bar s(l_0), \bar s(l_2), \bar s(l_4), \cdots \}$.
For this, we choose $l_{B_1}:l_A:l_{B_2} = 1:7:1$ and $l_b = l_{2m}$, hence $l_{C} = l_{2m+2}$.
Let $Q_{2m}$ denotes $Q(\alpha_0)$ in this setting.
Note $l_{2m}=9^{m}l_{0}$. 
We have
\begin{equation}
\begin{split}
	& Q_{2m} \le \frac{ 7\xi }{ \alpha_{0} } \bar s( 7 l_{2m} ) 
	         \le \frac{ 7\xi }{ \alpha_{0} } \bar s( l_{2m} ), \\
	& \bar s( l_{2m+2} ) \le \frac{ 1 }{ 1-Q_{2m} } \frac{2}{9} \bar s( l_{2m} ),
	\label{eq:101}
\end{split}
\end{equation}
where the first inequality uses Lemma~\ref{lem:3} and the subadditivity~\eqref{eq:sub}, and the second inequality uses the inequality~\eqref{eq:20} originated from Lemma~\ref{lem:4}.
We thus obtain recursive inequalities with the initial condition $\bar s(l_{0}) < \frac{ \alpha_{0} }{ 27\xi }$. 
The first values are $Q_0 < 7/27$ and $\bar s(l_2) < \alpha_0 / 90\xi$.
Then, for $m>0$, $\bar s(l_{2m})^2 l_{2m}$ decreases exponentially in $m$ as 
\begin{equation*}
	\frac{\bar s(l_{2m+2})^2 l_{2m+2}}{\bar s(l_{2m})^2 l_{2m}}
	< \frac{\bar s(l_{2})^2 l_{2}}{\bar s(l_{0})^2 l_{0}}
	= \frac{81}{100}.
\end{equation*}
As stressed in the previous section, this exponential decay is an important point that results in the convergence of the summation in Lemma~\ref{lem:7}.
We numerically obtain the upper-bounds of $\{\bar s(l_0), \bar s(l_2), \bar s(l_4), \cdots \}$ for their use in the following calculation.

Consider the summation $\sum_{m=0}^\infty \lambda_{2m}$.
In order to use Lemma~\ref{lem:6}, we choose $\alpha = \alpha_1 = \alpha_0/2 \in [1/3,1/2)$.
From Lemma~\ref{lem:3} and the inequality~\eqref{eq:92258} in Lemma~\ref{lem:6}, we obtain
\begin{equation}
	\begin{split}
		& \eta(2l_{2m},l_{2m},3l_{2m}) \\
		& \le 2 \frac{\xi}{2\alpha_1}\bar s(l_{2m})S(6l_{2m})-4 \frac{\xi}{2\alpha_1}\bar s(l_{2m}) \log \lll \frac{\xi}{2\alpha_1}\bar s(l_{2m}) \rrr \\
		& \quad + 2\epsilon(l_C,l_b,2\alpha_1) + \epsilon(l_B+l_b, l_b, 2\alpha_1).
	\end{split}
\end{equation}
As $\epsilon(\cdots)$ is much smaller than $(\xi/\alpha_1)\bar s(l_{2m})\epsilon_h$ and $\bar s(6l_{2m}) \le \bar s(l_{2m})$ from the subadditivity~\eqref{eq:sub},
\begin{equation}
	\begin{split}
		& \eta(2l_{2m},l_{2m},3l_{2m}) \\
		& \le \frac{12\xi}{\alpha_0}\bar s(l_{2m})^2 l_{2m}-\frac{4\xi}{\alpha_0}\bar s(l_{2m}) \log \lll \frac{\xi}{\alpha_0}\bar s(l_{2m}) \rrr.
	\end{split}
	\label{eq:50}
\end{equation}
Note that $\alpha_1$ was replaced by $\alpha_0/2$.
Similarly, 
\begin{equation}
	\begin{split}
		& \eta(l_{2m},l_{2m},l_{2m}) \\
		& \le \frac{12\xi}{\alpha_0}\bar s(l_{2m})^2 l_{2m}- \frac{8\xi}{\alpha_0}\bar s(l_{2m}) \log \lll \frac{2\xi}{\alpha_0}\bar s(l_{2m}) \rrr.
	\end{split}
	\label{eq:51}
\end{equation}
Along with the upper-bounds of $\{\bar s(l_{2m})\}$ obtained above, $\sum_{m=0}^\infty \lambda_{2m}$ can be obtained numerically.

The summation $\sum_{m=0}^\infty \lambda_{2m+1}$ is obtained similarly.
Here we frequently use the subadditivity, e.g., $\bar s(l_{2m+1}) \le \bar s(l_{2m})$, and $3l_{2m+1} = l_{2m+2}$.
We then obtain the following bounds similar to the above ones:
\begin{equation}
	\begin{split}
		& \eta(2l_{2m+1},l_{2m+1},3l_{2m+1}) \\
		& \le \frac{4\xi}{\alpha_0}\bar s(l_{2m}) \bar s(l_{2m+2}) l_{2m+2}-\frac{4\xi}{\alpha_0}\bar s(l_{2m}) \log \lll \frac{\xi}{\alpha_0}\bar s(l_{2m}) \rrr, \\
		& \eta(l_{2m+1},l_{2m+1},l_{2m+1}) \\
		& \le \frac{4\xi}{\alpha_0}\bar s(l_{2m}) \bar s(l_{2m+2}) l_{2m+2} - \frac{8\xi}{\alpha_0}\bar s(l_{2m}) \log \lll \frac{2\xi}{\alpha_0}\bar s(l_{2m}) \rrr.
	\end{split}
	\label{eq:52}
\end{equation}
From the set of upper-bounds of $\{\bar s(l_0), \bar s(l_2), \bar s(l_4), \cdots\}$ and the inequalities~\eqref{eq:50}, \eqref{eq:51}, and \eqref{eq:52}, we obtain the following bound:
\begin{equation}
	\sum_{i=0}^\infty \lambda_i < 0.1513\frac{\alpha_0}{\xi}l_0 + 5.893.
\end{equation}
By incorporating this and Lemma~\ref{lem:9} into Lemma~\ref{lem:7} and noting that $\eta(l_n, l_n, 2l_n)$ is exponentially small in $n$, we arrive at the following lemma:

\LEMMAb
	Inherit the definitions in Lemma~\ref{lem:9}.
	For any $\alpha_{0} \in [2/3,1)$, there exists integer $N$ such that for $n\ge N$,
	\begin{equation}
		S( 4\cdot 3^n l_0 ) < \frac{ \alpha_{0} }{ 2(1-\alpha_{0}) }
				  \( \log \frac{ \xi }{ 1 - \alpha_{0} } + 3 \)
				  4^{ n_0   } + 6.
	\end{equation}
	\label{lem:10}
\LEMMAe

\section{Discussion} \label{sec:discussion}

The essential ingredient of the proof was defining the entropy per site $\bar s(l)$ and the cut-off index $q(\alpha)$ associated with the cut-off proportion $Q(\alpha)$.
We obtained various relations of $\bar s(l)$ in terms of $Q(\alpha)$ for the partitioning in Fig.~\ref{fig:1}.
An important observation was that as the length scale is increased as $\ell_0 \rightarrow x \ell_0 \rightarrow x^2 \ell_0 \rightarrow \cdots \rightarrow x^n \ell_0$ with some positive integer $x$, $\bar s(x^n \ell_0)$ asymptotically decays exponentially in $n$
and this is accompanied by the asymptotic exponential decay of mutual information $I(A:C)$ in $n$ 
when $\{l_{B_1}, l_A, l_{B_2}\}$ are all comparable to $x^n \ell_0$.
This property motivated us to devise a renormalization-like construction in Fig.~\ref{fig:2} revealing that the entropy of a large region is equivalent to the entropy of the fixed end regions up to the correction terms composed of the mutual informations with different length scales.
The correction terms are upper-bounded by a constant thanks to the asymptotic decay of the mutual information.

Such an asymptotic behavior of $I(A:C)$ is, however, preceded by a transient stage in which $I(A:C)$ increases indefinitely.
During the transient stage, the decrement of $\bar s(x^n \ell_0)$ is slower and $Q(\alpha)$ is unbounded (note that in Lemma~\ref{lem:3}, $Q(\alpha)$ can be bounded only when $\bar s(l_A) < \alpha l_b/\xi l_A=\mathcal{O}[1/\xi]$).
This transient stage persists until the length scale reaches a certain value $l_0$ that is exponentially large in $\xi$.
Finding such $l_0$ was the first step of the previous section.
This transient behavior is responsible for the area-law bound being exponentially large in $\xi$.

One can slightly modify the logic towards Lemma~\ref{lem:6} in order to find the upper-bound of $I(B_1:B_2)$ instead.
That is, we write the state as
\begin{equation*}
	|\Psi\> = \sum_{i=1}^{M'}\sqrt{p'_i}|i'\>^{B_1}\otimes |\phi_i'\>^{AB_2C}
\end{equation*}
instead of Eq.~\eqref{eq:state}
and define  
\begin{equation*}
	Q'(\alpha) = \sum_{p'_i<2^{-\alpha l_A/\xi}} p'_i
\end{equation*}
instead of Eq.~\eqref{eq:Q}.
Then, we can obtain an upper-bound similar to that in Lemma~\ref{lem:6}:
\begin{equation*}
	\begin{split}
	I(B_1:B_2)
	& \le 2Q'(\alpha)S(B_2)-4Q'(\alpha)\log Q'(\alpha) \\
	& \quad + 2\epsilon(l_{B_2},l_A,\alpha) + \epsilon(l_B,l_A,2\alpha).
	\end{split}
\end{equation*}

The behavior of $I(A:C)$ and $I(B_1:B_2)$---the initial increment, saturation, and the asymptotic decay---is a characteristic feature of the states with exponentially decaying correlations, although the initial growth of the mutual information may be possibly absent.
There would be various kinds of states with different behaviors of the mutual information.
It is an interesting point that a state obeys an entanglement area law even when $I(A:C)$ asymptotically decays polynomially as $\mathcal{O}[n^{-k}]$ with $k>1$.
It seems that there is a room between the exponential and the polynomial decay, so the assumption~\eqref{eq:assum} might be slightly mitigable, although we do not have a further result.
This viewpoint suggests that the behavior of mutual information $I(A:C)$ in the IR limit is an important attribute governing the area-law scaling of entanglement entropies.

We note that the main assumption~\eqref{eq:assum} can be generalized to some extent.
Let $l_R = \min\{|R_X|,|R_Y|\}$ be the minimum of the sizes of $R_X$ and $R_Y$.
Then, one can replace the assumption by
\begin{equation}
\l| \<X \otimes Y\> - \<X\>\<Y\> \r| \le \text{Poly}(l,l_R) \|X\| \, \|Y\| 2^{-l/\xi},
\end{equation}
where $\text{Poly}(l,l_R)$ represents an arbitrary polynomial of $l$ and $l_R$.
This change basically alters the form of $\epsilon(L,l_b,\alpha)$ in Eq.~\eqref{eq:epsilon}, and hence the unit length $\ell_0$ in Eq.~(51).
Otherwise, the structure of the proof is left unchanged because throughout the proof, $l_R=\min\{l_A,l_C\}=l_A$ is always kept comparable to $l_b$ and thus the asymptotic behavior of $\epsilon(L,l_b,\alpha)$ is unchanged.

As a final remark, we note that the idea of taking a cut-off index $q(\alpha)$ is in line with the idea of approximating one-dimensional many-body states with matrix product states~\cite{sch11}.
In the language of the matrix product state, $q(\alpha)$ plays the role of a bond dimension, while $Q(\alpha)$ governs the accuracy of the approximation. 

\section{Acknowledgements}

This research was supported (in part) by the R\&D Convergence Program of NST (National Research Council of Science and Technology) of Republic of Korea (Grant No. CAP-15-08-KRISS). 

\section{Appendix}

\textbf{Proof of Lemma~\ref{lem:6}.}
i) If $q(\alpha)=0$, hence $Q(\alpha)=1$, the lemma follows from the triangle inequality $S(A)-S(C)\le S(AC)$.

ii) Suppose $0 < q(\alpha) < M$, hence $0 < Q(\alpha) < 1$.
We can split $\rho^{AC}$ into three parts:
\begin{equation}
	\begin{split}
	\rho^{AC} 
	& = \sum_{i,j=1}^M \sqrt{p_i p_j} |i\>^A\<j| \otimes \phi_{ij}^C \\
	& = \l[ \sum_{i,j=1}^{q(\alpha)} + \sum_{i,j=q(\alpha)+1}^M + \sum_{\text{the rest}} \r] \sqrt{p_i p_j} |i\>^A\<j| \otimes \phi_{ij}^C.
	\end{split}
	\label{eq:100002}
\end{equation}
Our aim here is to find the lower-bound of $S(\rho^{AC})$.

Let us first deal with the last sum.
Suppose we add to the system a single qubit $a$ initialized in state $|0\>^a$ and apply a local unitary transformation on $a+A$ such that $|0\>^a |i\>^A \rightarrow |0\>^a |i\>^A$ for $i \le q(\alpha)$ and $|0\>^a |i\>^A \rightarrow |1\>^a |i\>^A$ for $i > q(\alpha)$.
The resulting state is
\begin{equation*}
    |\tilde\Psi\> = \sqrt{1 - Q(\alpha)} |0\>^a |\Psi_{1q(\alpha)}\> 
                    + \sqrt{Q(\alpha)} |1\>^a |\Psi_{q(\alpha)+1,M}\>,
\end{equation*}
where $|\Psi_{mn}\> \equiv (P_{mn}/\sqrt{Q_{mn}})|\Psi\>$.
Letting $\tilde\rho = |\tilde\Psi\>\<\tilde\Psi|$,
\begin{equation}
	\tilde\rho^{AC} 
	= \l[ \sum_{i,j=1}^{q(\alpha)} + \sum_{i,j=q(\alpha)+1}^M \r] \sqrt{p_i p_j} |i\>^A\<j| \otimes \phi_{ij}^C,
\end{equation}
which differs from Eq.~\eqref{eq:100002} by the last sum.
Note that $|0\>^a\<0|\otimes\rho^{AC}$ and $\tilde\rho^{aAC}$ can be transformed to each other by a local unitary transformation on $a+A$, which implies $S(\rho^{AC})=S(\tilde\rho^{aAC})$.
Thus, from the triangle inequality,
\begin{equation}
	S(AC) = S(\tilde\rho^{aAC}) \ge S(\tilde\rho^{AC}) - S(\tilde\rho^a).
	\label{eq:54}
\end{equation}
Note $S(\tilde\rho^a)=H(\{Q(\alpha),1-Q(\alpha)\})$.

Let us now deal with the other two sums.
Introduce another state
\begin{equation}
	\begin{split}
    \tilde\sigma^{AC} 
    & = \sum_{i=1}^{q(\alpha)} p_i |i\>^A\<i| \otimes \phi_{ii}^C \\
    & \quad + \sum_{i,j=q(\alpha)+1}^M \sqrt{p_i p_j} |i\>^A\<j| \otimes \phi_{ij}^C,
    \end{split}
    \label{eq:62}
\end{equation}
which differs from $\tilde\rho^{AC}$ by the off-diagonal terms in the first sum. We find
\begin{equation}
\begin{split}
    & \mathcal{D} ( \tilde\rho^{AC}, \tilde\sigma^{AC} ) \\
    & = \frac{1}{2} \l\| \sum_{1\le i<j\le q(\alpha)} \sqrt{p_i p_j}
                         \( |i\>^A\<j| \otimes \phi_{ij}^C 
                           + |j\>^A\<i| \otimes \phi_{ji}^C \) \r\|_1 \\
    & \le \frac{1}{2} \sum_{1\le i<j\le q(\alpha)} \sqrt{p_i p_j}
                      \l\| |i\>^A\<j| \otimes \phi_{ij}^C 
                         + |j\>^A\<i| \otimes \phi_{ji}^C \r\|_1 .
\end{split}
\label{eq:56}
\end{equation}
Note
\begin{equation*}
\begin{split}
    & \l\| |i\>^A\<j| \otimes \phi_{ij}^C 
         + |j\>^A\<i| \otimes \phi_{ji}^C \r\|_1 \\
    & = \l\| X_{ij}^A \otimes   \( \phi_{ij}^C + \phi_{ji}^C \)
           + Y_{ij}^A \otimes i \( \phi_{ij}^C - \phi_{ji}^C \) \r\|_1 \\
    & \le 2 \lll \l\| \phi_{ij}^C + \phi_{ji}^C \r\|_1 
               + \l\| i \( \phi_{ij}^C - \phi_{ji}^C \) \r\|_1 \rrr,
\end{split}
\end{equation*}
where $X_{ij}^A = |i\>^A\<j| + |j\>^A\<i|$ and $Y_{ij}^A = -i |i\>^A\<j| + i |j\>^A\<i|$ are the Pauli matrices. 
As the term inside the norm is hermitian having real eigenvalues, one can write, e.g., the first term as
\begin{equation*}
    \l\| \phi_{ij}^C + \phi_{ji}^C \r\|_1
    = \tr\lll \( \phi_{ij}^C + \phi_{ji}^C \) \Lambda_{ij}^C \rrr
\end{equation*}
for some matrix $\Lambda_{ij}^C$ that has $\pm 1$ as eigenvalues. 
Thus,
\begin{equation*}
	\begin{split}
    \sqrt{p_i p_j} \l\| \phi_{ij}^C + \phi_{ji}^C \r\|_1
    & =  \l\< X_{ij}^A \otimes \Lambda_{ij}^C \r\>  \\
    & \le  \l\< X_{ij}^A \r\> \l\< \Lambda_{ij}^C \r\>
          + \l\| X_{ij}^A \r\| \l\| \Lambda_{ij}^C \r\|
            2^{-l_b / \xi}  \\
    & =  2^{-l_b / \xi},
    \end{split}
\end{equation*}
where we use the Schmidt decomposition~\eqref{eq:state} in the first line, 
the assumption~\eqref{eq:assum} in the second line, 
and $\<X_{ij}^A\> = 0$ and $\l\| X_{ij}^A \r\| = \l\| \Lambda_{ij}^C \r\| = 1$ in the last line. 
The other norm is bounded in the same manner. We thus find
\begin{equation*}
	\begin{split}
    \mathcal{D} \( \tilde\rho^{AC}, \tilde\sigma^{AC} \) 
    & \le  \text{(\# of terms in Eq.~\eqref{eq:56})} 
          \cdot 2 \cdot 2^{-l_b / \xi} \\
    & \le  2 \frac{q(\alpha)\{q(\alpha)-1\}}{2} 2^{-l_b / \xi}
    \le 2^{-(1-2\alpha) l_b / \xi},
    \end{split}
\end{equation*}
where we use $q(\alpha) \le 2^{\alpha l_b / \xi}$ (otherwise $\sum_{i=1}^{q(\alpha)} p_i > 1$).
Note that $\tilde\sigma^{AC}$ is purified by attaching a system with Hilbert-space dimension $q(\alpha)+1$ and the region $B$, while $\tilde\rho^{AC}$ is purified by attaching a qubit and the region $B$ (i.e., $|\tilde\Psi\>$ is the purification), which means $\tilde\sigma^{AC}$ has a larger Hilbert-space dimension.
Furthermore, both the states share the same basis states. 
Consequently, the Fannes' inequality~\eqref{eq:fineq} can be applied with $d = \{ q(\alpha) + 1 \} 2^{l_B}$.
From $d \le ( 2^{\alpha l_b / \xi} + 1 ) 2^{l_B} < 2^{l_b + l_B}$, it follows
\begin{equation}
    \l| S ( \tilde\rho^{AC} ) - S ( \tilde\sigma^{AC} ) \r| \\
    \le \epsilon(l_B + l_b, l_b, 2\alpha).
    \label{eq:57}
\end{equation}

We are now in a position to lower-bound $S(AC)$. 
For convenience, let $\tilde \sigma_{mn}^{AC} \equiv (1/Q_{mn})P_{mn}\tilde\sigma^{AC}P_{mn}$.
Then, Eq.~\eqref{eq:62} can be written as
\begin{equation}
	\tilde\sigma^{AC} = \{1 - Q(\alpha)\} \tilde\sigma_{1q(\alpha)}^{AC} 
	+ Q(\alpha) \tilde\sigma_{q(\alpha)+1,M}^{AC}.
	\label{eq:58}
\end{equation}
Note
\begin{widetext}
\begin{equation}
\begin{split}
    \{1 -Q(\alpha)\} S ( \tilde\sigma_{1q(\alpha)}^{AC} ) 
    & = - \{1-Q(\alpha)\} \lll \sum_{i=1}^{q(\alpha)} \frac{p_i}{1-Q(\alpha)} \log \frac{p_i}{1-Q(\alpha)} + \sum_{i=1}^{q(\alpha)} \frac{p_i}{1-Q(\alpha)} S ( \phi_{ii}^C ) \rrr \\
    & = \sum_{i=1}^{q(\alpha)} p_i \lll -\log p_i + S ( \phi_{ii}^C ) \rrr 
        + \{1-Q(\alpha)\} \log \{1-Q(\alpha)\} \\
    & \ge - \sum_{i=1}^{q(\alpha)} p_i \log p_i 
          + \{ 1-Q(\alpha) \} \{ S(C) - \epsilon(l_C, l_b, \alpha) \} 
          + \{ 1-Q(\alpha) \} \log \{ 1-Q(\alpha) \},
\end{split}
\label{eq:59}
\end{equation}
where the first line comes from a general equality $S( \sum_i p_i |i\>\<i| \otimes \rho_i ) = -\sum_i p_i \log p_i + \sum_i p_i S( \rho_i )$ satisfied when $\{|i\>\}$ is an orthonormal basis~\cite{nie00} and 
the third line from Lemma~\ref{lem:2}. 
Note also
\begin{equation}
\begin{split}
    Q(\alpha) S ( \tilde\sigma_{q(\alpha)+1,M}^{AC} ) 
    & \ge Q(\alpha) S ( \tilde\sigma_{q(\alpha)+1,M}^{A} )
          - Q(\alpha) S ( \tilde\sigma_{q(\alpha)+1,M}^{C} ) \\
    & \ge - Q(\alpha) \sum_{i=q(\alpha)+1}^M \frac{p_i}{Q(\alpha)} \log \frac{p_i}{Q(\alpha)}  
          - \llll S ( \tilde\sigma^C ) - \{ 1 - Q(\alpha) \} 
                  S ( \tilde\sigma_{1q(\alpha)}^{C} ) \rrrr \\
    & = - \sum_{i=q(\alpha)+1}^M p_i \log p_i + Q(\alpha) \log Q(\alpha) 
        - Q(\alpha) S(C) - \{ 1-Q(\alpha) \} \epsilon(l_C, l_b, \alpha),
\end{split}
\label{eq:60}
\end{equation}
where the first line comes from the triangle inequality of entropy,
the second line from the concavity of entropy,
and the third line from Lemma~\ref{lem:2} and $S(\tilde\sigma^C) = S(C)$.
From the inequalities~\eqref{eq:54} and \eqref{eq:57}, we have
\begin{equation}
	S(AC) \ge S(\tilde\sigma^{AC}) - S(\tilde\rho^a) - \epsilon(l_B+l_b, l_b, 2\alpha).
	\label{eq:61}
\end{equation}
Moreover, from Eqs.~\eqref{eq:58}, \eqref{eq:59}, \eqref{eq:60}, and the concavity of entropy, 
\begin{equation}
	S(\tilde\sigma^{AC}) \ge S(A) + S(C) - 2Q(\alpha)S(C)  
	- H(\{Q(\alpha),1-Q(\alpha)\}) - 2\epsilon(l_C, l_b, \alpha).
\end{equation}
Combining these two inequalities and noting $S( \tilde\rho^a ) = H(\{Q(\alpha),1-Q(\alpha)\})\le -2 Q(\alpha) \log Q(\alpha)$ for $Q(\alpha) \le 1/2$, the lemma follows.
\end{widetext}

iii) If $q(\alpha) = M$, hence $Q(\alpha) = 0$, the extra qubit $a$ is not necessary any more and the inequality~\eqref{eq:61} becomes
$S(AC) \ge S(\tilde\sigma^{AC}) - \epsilon(l_B+l_b, l_b, 2\alpha)$.
Then, the lemma again follows from the inequality~\eqref{eq:59} with $Q(\alpha) = 0$.
\qed

\bibliography{references}

\begin{thebibliography}{24}%
\makeatletter
\providecommand \@ifxundefined [1]{%
 \@ifx{#1\undefined}
}%
\providecommand \@ifnum [1]{%
 \ifnum #1\expandafter \@firstoftwo
 \else \expandafter \@secondoftwo
 \fi
}%
\providecommand \@ifx [1]{%
 \ifx #1\expandafter \@firstoftwo
 \else \expandafter \@secondoftwo
 \fi
}%
\providecommand \natexlab [1]{#1}%
\providecommand \enquote  [1]{``#1''}%
\providecommand \bibnamefont  [1]{#1}%
\providecommand \bibfnamefont [1]{#1}%
\providecommand \citenamefont [1]{#1}%
\providecommand \href@noop [0]{\@secondoftwo}%
\providecommand \href [0]{\begingroup \@sanitize@url \@href}%
\providecommand \@href[1]{\@@startlink{#1}\@@href}%
\providecommand \@@href[1]{\endgroup#1\@@endlink}%
\providecommand \@sanitize@url [0]{\catcode `\\12\catcode `\$12\catcode
  `\&12\catcode `\#12\catcode `\^12\catcode `\_12\catcode `\%12\relax}%
\providecommand \@@startlink[1]{}%
\providecommand \@@endlink[0]{}%
\providecommand \url  [0]{\begingroup\@sanitize@url \@url }%
\providecommand \@url [1]{\endgroup\@href {#1}{\urlprefix }}%
\providecommand \urlprefix  [0]{URL }%
\providecommand \Eprint [0]{\href }%
\providecommand \doibase [0]{http://dx.doi.org/}%
\providecommand \selectlanguage [0]{\@gobble}%
\providecommand \bibinfo  [0]{\@secondoftwo}%
\providecommand \bibfield  [0]{\@secondoftwo}%
\providecommand \translation [1]{[#1]}%
\providecommand \BibitemOpen [0]{}%
\providecommand \bibitemStop [0]{}%
\providecommand \bibitemNoStop [0]{.\EOS\space}%
\providecommand \EOS [0]{\spacefactor3000\relax}%
\providecommand \BibitemShut  [1]{\csname bibitem#1\endcsname}%
\let\auto@bib@innerbib\@empty
\bibitem [{\citenamefont {Eisert}\ \emph {et~al.}(2010)\citenamefont {Eisert},
  \citenamefont {Cramer},\ and\ \citenamefont {Plenio}}]{eis10}%
  \BibitemOpen
  \bibfield  {author} {\bibinfo {author} {\bibfnamefont {J.}~\bibnamefont
  {Eisert}}, \bibinfo {author} {\bibfnamefont {M.}~\bibnamefont {Cramer}}, \
  and\ \bibinfo {author} {\bibfnamefont {M.~B.}\ \bibnamefont {Plenio}},\
  }\bibfield  {title} {\enquote {\bibinfo {title} {Colloquium: {Area} {Laws}
  for the {Entanglement} {Entropy}},}\ }\href@noop {} {\bibfield  {journal}
  {\bibinfo  {journal} {Rev. Mod. Phys.}\ }\textbf {\bibinfo {volume} {82}},\
  \bibinfo {pages} {277} (\bibinfo {year} {2010})}\BibitemShut {NoStop}%
\bibitem [{\citenamefont {Hastings}\ and\ \citenamefont {Koma}(2006)}]{has06}%
  \BibitemOpen
  \bibfield  {author} {\bibinfo {author} {\bibfnamefont {M.~B.}\ \bibnamefont
  {Hastings}}\ and\ \bibinfo {author} {\bibfnamefont {T.}~\bibnamefont
  {Koma}},\ }\bibfield  {title} {\enquote {\bibinfo {title} {Spectral {Gap} and
  {Exponential} {Decay} of {Correlations}},}\ }\href@noop {} {\bibfield
  {journal} {\bibinfo  {journal} {Commun. Math. Phys.}\ }\textbf {\bibinfo
  {volume} {265}},\ \bibinfo {pages} {781} (\bibinfo {year}
  {2006})}\BibitemShut {NoStop}%
\bibitem [{\citenamefont {Nachtergaele}\ and\ \citenamefont
  {Sims}(2006)}]{nac06}%
  \BibitemOpen
  \bibfield  {author} {\bibinfo {author} {\bibfnamefont {B.}~\bibnamefont
  {Nachtergaele}}\ and\ \bibinfo {author} {\bibfnamefont {R.}~\bibnamefont
  {Sims}},\ }\bibfield  {title} {\enquote {\bibinfo {title} {Lieb-{Robinson}
  {Bounds} and the {Exponential} {Clustering} {Theorem}},}\ }\href@noop {}
  {\bibfield  {journal} {\bibinfo  {journal} {Commun. Math. Phys.}\ }\textbf
  {\bibinfo {volume} {265}},\ \bibinfo {pages} {119} (\bibinfo {year}
  {2006})}\BibitemShut {NoStop}%
\bibitem [{\citenamefont {Gosset}\ and\ \citenamefont {Huang}(2016)}]{gos16}%
  \BibitemOpen
  \bibfield  {author} {\bibinfo {author} {\bibfnamefont {D.}~\bibnamefont
  {Gosset}}\ and\ \bibinfo {author} {\bibfnamefont {Y.}~\bibnamefont {Huang}},\
  }\bibfield  {title} {\enquote {\bibinfo {title} {Correlation {Length} versus
  {Gap} in {Frustration}-{Free} {Systems}},}\ }\href@noop {} {\bibfield
  {journal} {\bibinfo  {journal} {Phys. Rev. Lett.}\ }\textbf {\bibinfo
  {volume} {116}},\ \bibinfo {pages} {097202} (\bibinfo {year}
  {2016})}\BibitemShut {NoStop}%
\bibitem [{\citenamefont {Hastings}()}]{has07}%
  \BibitemOpen
  \bibfield  {author} {\bibinfo {author} {\bibfnamefont {M.~B.}\ \bibnamefont
  {Hastings}},\ }\bibfield  {title} {\enquote {\bibinfo {title} {An {Area}
  {Law} for {One}-{Dimensional} {Quantum} {Systems}},}\ }\href@noop {}
  {\bibinfo  {journal} {J. Stat. Mech. (2007) P08024}\ }\BibitemShut {NoStop}%
\bibitem [{\citenamefont {Aharonov}\ \emph {et~al.}(2011)\citenamefont
  {Aharonov}, \citenamefont {Arad}, \citenamefont {Vazirani},\ and\
  \citenamefont {Landau}}]{aha11}%
  \BibitemOpen
\bibfield  {journal} {  }\bibfield  {author} {\bibinfo {author} {\bibfnamefont
  {D.}~\bibnamefont {Aharonov}}, \bibinfo {author} {\bibfnamefont
  {I.}~\bibnamefont {Arad}}, \bibinfo {author} {\bibfnamefont {U.}~\bibnamefont
  {Vazirani}}, \ and\ \bibinfo {author} {\bibfnamefont {Z.}~\bibnamefont
  {Landau}},\ }\bibfield  {title} {\enquote {\bibinfo {title} {The
  {Detectability} {Lemma} and {Its} {Applications} to {Quantum} {Hamiltonian}
  {Complexity}},}\ }\href@noop {} {\bibfield  {journal} {\bibinfo  {journal}
  {New J. Phys.}\ }\textbf {\bibinfo {volume} {13}},\ \bibinfo {pages} {113043}
  (\bibinfo {year} {2011})}\BibitemShut {NoStop}%
\bibitem [{\citenamefont {Arad}\ \emph {et~al.}(2012)\citenamefont {Arad},
  \citenamefont {Landau},\ and\ \citenamefont {Vazirani}}]{ara12}%
  \BibitemOpen
  \bibfield  {author} {\bibinfo {author} {\bibfnamefont {I.}~\bibnamefont
  {Arad}}, \bibinfo {author} {\bibfnamefont {Z.}~\bibnamefont {Landau}}, \ and\
  \bibinfo {author} {\bibfnamefont {U.}~\bibnamefont {Vazirani}},\ }\bibfield
  {title} {\enquote {\bibinfo {title} {Improved {One}-{Dimensional} {Area}
  {Law} for {Frustration}-{Free} {Systems}},}\ }\href@noop {} {\bibfield
  {journal} {\bibinfo  {journal} {Phys. Rev. B}\ }\textbf {\bibinfo {volume}
  {85}},\ \bibinfo {pages} {195145} (\bibinfo {year} {2012})}\BibitemShut
  {NoStop}%
\bibitem [{\citenamefont {Arad}\ \emph {et~al.}()\citenamefont {Arad},
  \citenamefont {Kitaev}, \citenamefont {Landau},\ and\ \citenamefont
  {Vazirani}}]{ara13}%
  \BibitemOpen
  \bibfield  {author} {\bibinfo {author} {\bibfnamefont {I.}~\bibnamefont
  {Arad}}, \bibinfo {author} {\bibfnamefont {A.}~\bibnamefont {Kitaev}},
  \bibinfo {author} {\bibfnamefont {Z.}~\bibnamefont {Landau}}, \ and\ \bibinfo
  {author} {\bibfnamefont {U.}~\bibnamefont {Vazirani}},\ }\href@noop {}
  {\enquote {\bibinfo {title} {An {Area} {Law} and {Sub}-{Exponential}
  {Algorithm} for {1D} {Systems}},}\ }\Eprint
  {http://arxiv.org/abs/arXiv:1301.1162} {arXiv:1301.1162} \BibitemShut
  {NoStop}%
\bibitem [{\citenamefont {de~Beaudrap}\ \emph {et~al.}(2010)\citenamefont
  {de~Beaudrap}, \citenamefont {Ohliger}, \citenamefont {Osborne},\ and\
  \citenamefont {Eisert}}]{deb10}%
  \BibitemOpen
  \bibfield  {author} {\bibinfo {author} {\bibfnamefont {N.}~\bibnamefont
  {de~Beaudrap}}, \bibinfo {author} {\bibfnamefont {M.}~\bibnamefont
  {Ohliger}}, \bibinfo {author} {\bibfnamefont {T.~J.}\ \bibnamefont
  {Osborne}}, \ and\ \bibinfo {author} {\bibfnamefont {J.}~\bibnamefont
  {Eisert}},\ }\bibfield  {title} {\enquote {\bibinfo {title} {Solving
  {Frustration}-{Free} {Spin} {Systems}},}\ }\href@noop {} {\bibfield
  {journal} {\bibinfo  {journal} {Phys. Rev. Lett.}\ }\textbf {\bibinfo
  {volume} {105}},\ \bibinfo {pages} {060504} (\bibinfo {year}
  {2010})}\BibitemShut {NoStop}%
\bibitem [{\citenamefont {Cho}(2014)}]{cho14}%
  \BibitemOpen
  \bibfield  {author} {\bibinfo {author} {\bibfnamefont {J.}~\bibnamefont
  {Cho}},\ }\bibfield  {title} {\enquote {\bibinfo {title} {Sufficient
  {Condition} for {Entanglement} {Area} {Laws} in {Thermodynamically} {Gapped}
  {Spin} {Systems}},}\ }\href@noop {} {\bibfield  {journal} {\bibinfo
  {journal} {Phys. Rev. Lett.}\ }\textbf {\bibinfo {volume} {113}},\ \bibinfo
  {pages} {197204} (\bibinfo {year} {2014})}\BibitemShut {NoStop}%
\bibitem [{\citenamefont {Hastings}(2007)}]{has07b}%
  \BibitemOpen
  \bibfield  {author} {\bibinfo {author} {\bibfnamefont {M.~B.}\ \bibnamefont
  {Hastings}},\ }\bibfield  {title} {\enquote {\bibinfo {title} {Entropy and
  {Entanglement} in {Quantum} {Ground} {States}},}\ }\href@noop {} {\bibfield
  {journal} {\bibinfo  {journal} {Phys. Rev. B}\ }\textbf {\bibinfo {volume}
  {76}},\ \bibinfo {pages} {035114} (\bibinfo {year} {2007})}\BibitemShut
  {NoStop}%
\bibitem [{\citenamefont {Brand\~ao}\ and\ \citenamefont
  {Horodecki}(2013)}]{bra13}%
  \BibitemOpen
  \bibfield  {author} {\bibinfo {author} {\bibfnamefont {F.~G. S.~L.}\
  \bibnamefont {Brand\~ao}}\ and\ \bibinfo {author} {\bibfnamefont
  {M.}~\bibnamefont {Horodecki}},\ }\bibfield  {title} {\enquote {\bibinfo
  {title} {An {Area} {Law} for {Entanglement} from {Exponential} {Decay} of
  {Correlations}},}\ }\href@noop {} {\bibfield  {journal} {\bibinfo  {journal}
  {Nature Physics}\ }\textbf {\bibinfo {volume} {9}},\ \bibinfo {pages} {721}
  (\bibinfo {year} {2013})}\BibitemShut {NoStop}%
\bibitem [{\citenamefont {Brand\~ao}\ and\ \citenamefont
  {Horodecki}(2014)}]{bra14}%
  \BibitemOpen
  \bibfield  {author} {\bibinfo {author} {\bibfnamefont {F.~G. S.~L.}\
  \bibnamefont {Brand\~ao}}\ and\ \bibinfo {author} {\bibfnamefont
  {M.}~\bibnamefont {Horodecki}},\ }\bibfield  {title} {\enquote {\bibinfo
  {title} {Exponential {Decay} of {Correlations} {Implies} {Area} {Law}},}\
  }\href@noop {} {\bibfield  {journal} {\bibinfo  {journal} {Commun. Math.
  Phys.}\ }\textbf {\bibinfo {volume} {333}},\ \bibinfo {pages} {761} (\bibinfo
  {year} {2014})}\BibitemShut {NoStop}%
\bibitem [{\citenamefont {Bekenstein}(1973)}]{bek73}%
  \BibitemOpen
  \bibfield  {author} {\bibinfo {author} {\bibfnamefont {J.~D.}\ \bibnamefont
  {Bekenstein}},\ }\bibfield  {title} {\enquote {\bibinfo {title} {Black
  {Holes} and {Entropy}},}\ }\href@noop {} {\bibfield  {journal} {\bibinfo
  {journal} {Phys. Rev. D}\ }\textbf {\bibinfo {volume} {7}},\ \bibinfo {pages}
  {2333} (\bibinfo {year} {1973})}\BibitemShut {NoStop}%
\bibitem [{\citenamefont {Hawking}(1974)}]{haw74}%
  \BibitemOpen
  \bibfield  {author} {\bibinfo {author} {\bibfnamefont {S.~W.}\ \bibnamefont
  {Hawking}},\ }\bibfield  {title} {\enquote {\bibinfo {title} {Black {Hole}
  {Explosions}?}}\ }\href@noop {} {\bibfield  {journal} {\bibinfo  {journal}
  {Nature}\ }\textbf {\bibinfo {volume} {248}},\ \bibinfo {pages} {30}
  (\bibinfo {year} {1974})}\BibitemShut {NoStop}%
\bibitem [{\citenamefont {Schollw\"ock}(2011)}]{sch11}%
  \BibitemOpen
  \bibfield  {author} {\bibinfo {author} {\bibfnamefont {U.}~\bibnamefont
  {Schollw\"ock}},\ }\bibfield  {title} {\enquote {\bibinfo {title} {The
  {Density}-{Matrix} {Renormalization} {Group} in the {Age} of {Matrix}
  {Product} {States}},}\ }\href@noop {} {\bibfield  {journal} {\bibinfo
  {journal} {Ann. Phys.}\ }\textbf {\bibinfo {volume} {326}},\ \bibinfo {pages}
  {96} (\bibinfo {year} {2011})}\BibitemShut {NoStop}%
\bibitem [{\citenamefont {Or\'us}(2014)}]{oru14}%
  \BibitemOpen
  \bibfield  {author} {\bibinfo {author} {\bibfnamefont {R.}~\bibnamefont
  {Or\'us}},\ }\bibfield  {title} {\enquote {\bibinfo {title} {A {Practical}
  {Introduction} to {Tensor} {Networks}: {Matrix} {Product} {States} and
  {Projected} {Entangled} {Pair} {States}},}\ }\href@noop {} {\bibfield
  {journal} {\bibinfo  {journal} {Ann. Phys.}\ }\textbf {\bibinfo {volume}
  {349}},\ \bibinfo {pages} {117} (\bibinfo {year} {2014})}\BibitemShut
  {NoStop}%
\bibitem [{\citenamefont {Kitaev}\ and\ \citenamefont
  {Preskill}(2006)}]{kit06}%
  \BibitemOpen
  \bibfield  {author} {\bibinfo {author} {\bibfnamefont {A.}~\bibnamefont
  {Kitaev}}\ and\ \bibinfo {author} {\bibfnamefont {J.}~\bibnamefont
  {Preskill}},\ }\bibfield  {title} {\enquote {\bibinfo {title} {Topological
  {Entanglement} {Entropy}},}\ }\href@noop {} {\bibfield  {journal} {\bibinfo
  {journal} {Phys. Rev. Lett.}\ }\textbf {\bibinfo {volume} {96}},\ \bibinfo
  {pages} {110404} (\bibinfo {year} {2006})}\BibitemShut {NoStop}%
\bibitem [{\citenamefont {Levin}\ and\ \citenamefont {Wen}(2006)}]{lev06}%
  \BibitemOpen
  \bibfield  {author} {\bibinfo {author} {\bibfnamefont {M.}~\bibnamefont
  {Levin}}\ and\ \bibinfo {author} {\bibfnamefont {X.-G.}\ \bibnamefont
  {Wen}},\ }\bibfield  {title} {\enquote {\bibinfo {title} {Detecting
  {Topological} {Order} in a {Ground} {State} {Wave} {Function}},}\ }\href@noop
  {} {\bibfield  {journal} {\bibinfo  {journal} {Phys. Rev. Lett.}\ }\textbf
  {\bibinfo {volume} {96}},\ \bibinfo {pages} {110405} (\bibinfo {year}
  {2006})}\BibitemShut {NoStop}%
\bibitem [{\citenamefont {Ryu}\ and\ \citenamefont {Takayanagi}(2006)}]{ryu06}%
  \BibitemOpen
  \bibfield  {author} {\bibinfo {author} {\bibfnamefont {S.}~\bibnamefont
  {Ryu}}\ and\ \bibinfo {author} {\bibfnamefont {T.}~\bibnamefont
  {Takayanagi}},\ }\bibfield  {title} {\enquote {\bibinfo {title} {Holographic
  {Derivation} of {Entanglement} {Entropy} from the anti--de {Sitter}
  {Space}/{Conformal} {Field} {Theory} {Correspondence}},}\ }\href@noop {}
  {\bibfield  {journal} {\bibinfo  {journal} {Phys. Rev. Lett.}\ }\textbf
  {\bibinfo {volume} {96}},\ \bibinfo {pages} {181602} (\bibinfo {year}
  {2006})}\BibitemShut {NoStop}%
\bibitem [{\citenamefont {Gharibian}\ \emph {et~al.}()\citenamefont
  {Gharibian}, \citenamefont {Huang}, \citenamefont {Landau},\ and\
  \citenamefont {Shin}}]{gha14}%
  \BibitemOpen
  \bibfield  {author} {\bibinfo {author} {\bibfnamefont {S.}~\bibnamefont
  {Gharibian}}, \bibinfo {author} {\bibfnamefont {Y.}~\bibnamefont {Huang}},
  \bibinfo {author} {\bibfnamefont {Z.}~\bibnamefont {Landau}}, \ and\ \bibinfo
  {author} {\bibfnamefont {S.~W.}\ \bibnamefont {Shin}},\ }\href@noop {}
  {\enquote {\bibinfo {title} {Quantum {Hamiltonian} {Complexity}},}\ }\Eprint
  {http://arxiv.org/abs/arXiv:1401.3916} {arXiv:1401.3916} \BibitemShut
  {NoStop}%
\bibitem [{\citenamefont {Hayden}\ \emph {et~al.}(2004)\citenamefont {Hayden},
  \citenamefont {Leung}, \citenamefont {Shor},\ and\ \citenamefont
  {Winter}}]{hay04}%
  \BibitemOpen
  \bibfield  {author} {\bibinfo {author} {\bibfnamefont {P.}~\bibnamefont
  {Hayden}}, \bibinfo {author} {\bibfnamefont {D.}~\bibnamefont {Leung}},
  \bibinfo {author} {\bibfnamefont {P.~W.}\ \bibnamefont {Shor}}, \ and\
  \bibinfo {author} {\bibfnamefont {A.}~\bibnamefont {Winter}},\ }\bibfield
  {title} {\enquote {\bibinfo {title} {Randomizing {Quantum} {States}:
  {Constructions} and {Applications}},}\ }\href@noop {} {\bibfield  {journal}
  {\bibinfo  {journal} {Comm. Math. Phys.}\ }\textbf {\bibinfo {volume}
  {250}},\ \bibinfo {pages} {371} (\bibinfo {year} {2004})}\BibitemShut
  {NoStop}%
\bibitem [{\citenamefont {Hastings}(2016)}]{has16}%
  \BibitemOpen
  \bibfield  {author} {\bibinfo {author} {\bibfnamefont {M.}~\bibnamefont
  {Hastings}},\ }\bibfield  {title} {\enquote {\bibinfo {title} {Random {MERA}
  {States} and the {Tightness} of the {Brandao}-{Horodecki} {Entropy}
  {Bound}},}\ }\href@noop {} {\bibfield  {journal} {\bibinfo  {journal}
  {Quantum Inf. Comput.}\ }\textbf {\bibinfo {volume} {16}},\ \bibinfo {pages}
  {1228} (\bibinfo {year} {2016})}\BibitemShut {NoStop}%
\bibitem [{\citenamefont {Nielsen}\ and\ \citenamefont {Chuang}(2000)}]{nie00}%
  \BibitemOpen
  \bibfield  {author} {\bibinfo {author} {\bibfnamefont {M.}~\bibnamefont
  {Nielsen}}\ and\ \bibinfo {author} {\bibfnamefont {I.}~\bibnamefont
  {Chuang}},\ }\href@noop {} {\emph {\bibinfo {title} {Quantum {Computation}
  and {Quantum} {Information}}}}\ (\bibinfo  {publisher} {Cambridge University
  Press, Cambridge, England},\ \bibinfo {year} {2000})\BibitemShut {NoStop}%
\end{thebibliography}%

\end{document}